\documentclass{Nature}
\usepackage{graphicx}
\usepackage{aas_macros,amsmath,amssymb,xcolor,url}
\usepackage{multirow}
\usepackage{longtable}
\usepackage{placeins}
\usepackage{lineno}
\usepackage[utf8]{inputenc}  
\usepackage[T1]{fontenc}
\usepackage{graphicx,psfrag}
\usepackage{mathrsfs}
\usepackage{amsmath,amsfonts,amssymb,pifont,gensymb}
\usepackage{multirow,enumerate}
\usepackage{comment,hyperref}
\usepackage{xcolor}
\usepackage{acronym}
\usepackage{xspace}
\usepackage[normalem]{ulem}
\usepackage{mathtools}
\usepackage{subfigure}
\usepackage{makecell}
\usepackage{diagbox}
\newcommand{\n}{_\mathrm{n}}
\newcommand{\p}{_\mathrm{p}}
\newcommand{\ns}{n_\mathrm{sat}}

\newcommand{\MeV}{\,\mathrm{MeV}}

\newcommand{\fmiq}{\,\mathrm{fm}^{-3}}

\newcommand{\esym}{S_0}%{E_\mathrm{sym}}
\newcommand{\gasy}{\gamma_\mathrm{asy}}

\title{Constraining Neutron-Star Matter with Microscopic and Macroscopic Collisions}

\author{% Group 1
        Sabrina Huth$^{1,2,*\dagger}$,
        Peter T.~H.~Pang$^{3,4*\dagger}$,
        Ingo Tews$^{5}$,
        Tim Dietrich$^{6,7}$,
        Arnaud Le F{\`e}vre$^{8}$,
        Achim Schwenk$^{1,2,9}$,
        Wolfgang Trautmann$^{8}$,
        Kshitij Agarwal$^{10}$, 
        Mattia Bulla$^{11}$,
        Michael W.~Coughlin$^{12}$,
        Chris Van Den Broeck$^{3,4}$
    }

\begin{document} 
\maketitle

\begin{affiliations}
\item Technische Universit{\"at} Darmstadt, Department of Physics, D-64289 Darmstadt, Germany
\item ExtreMe Matter Institute EMMI, GSI Helmholtzzentrum f\"ur Schwerionenforschung GmbH, D-64291 Darmstadt, Germany
\item Nikhef, Science Park 105, 1098 XG Amsterdam, The Netherlands
\item Institute for Gravitational and Subatomic Physics (GRASP), Utrecht University, Princetonplein 1, 3584 CC Utrecht, The Netherlands
\item Theoretical Division, Los Alamos National Laboratory, Los Alamos, NM 87545, USA
\item Institut f\"{u}r Physik und Astronomie, Universit\"{a}t Potsdam, D-14476 Potsdam, Germany
\item Max Planck Institute for Gravitational Physics (Albert Einstein Institute), Am M\"uhlenberg 1, D-14476 Potsdam, Germany
\item GSI Helmholtzzentrum f\"ur Schwerionenforschung GmbH, D-64291 Darmstadt, Germany
\item Max-Planck-Institut f\"ur Kernphysik, Saupfercheckweg 1, D-69117 Heidelberg, Germany
\item Physikalisches Institut, Eberhard Karls Universität Tübingen, D-72076 Tübingen, Germany 
\item The Oskar Klein Centre, Department of Astronomy, Stockholm University, AlbaNova, SE-10691 Stockholm, Sweden
\item School of Physics and Astronomy, University of Minnesota, Minneapolis, Minnesota 55455, USA
\end{affiliations}

\noindent $^*$ These authors contributed equally: Sabrina Huth, Peter T.H. Pang.\\
E-mail: shuth@theorie.ikp.physik.tu-darmstadt.de, t.h.pang@uu.nl\\

\begin{abstract}
Interpreting high-energy, astrophysical phenomena, such as supernova explosions or neutron-star collisions, requires a robust understanding of matter at supranuclear densities.
However, our knowledge about dense matter explored in the cores of neutron stars remains limited. 
Fortunately, dense matter is not probed only in astrophysical observations, but also in terrestrial heavy-ion collision experiments. 
Here we use Bayesian inference to combine data from astrophysical multi-messenger observations of neutron stars\cite{TheLIGOScientific:2017qsa,Abbott:2020uma,Monitor:2017mdv,Coughlin:2017ydf,Miller:2019cac,Riley:2019yda,Miller:2021qha,Riley:2021pdl,Dietrich:2020lps} and from heavy-ion collisions of gold nuclei at relativistic energies\cite{Fevre:2015fza,Russotto:2016ucm} with microscopic nuclear theory calculations\cite{Hebeler:2013nza,Tews:2012fj,Lynn:2016, Drischler:2017wtt, Dris20GPB, Huth_2021} to improve our understanding of dense matter. 
We find that the inclusion of heavy-ion collision data indicates an increase in the pressure in dense matter relative to previous analyses, shifting neutron-star radii towards larger values, consistent with recent observations by the Neutron Star Interior Composition Explorer mission\cite{Miller:2019cac,Riley:2019yda,Miller:2021qha,Riley:2021pdl,Raaijmakers:2021uju}.
Our findings show that constraints from heavy-ion collision experiments show a remarkable consistency with multi-messenger observations and provide complementary information on nuclear matter at intermediate densities.
This work combines nuclear theory, nuclear experiment and astrophysical observations, and shows how joint analyses can shed light on the properties of neutron-rich supranuclear matter over the density range probed in neutron stars.
\end{abstract}

%\keywords{nuclear physics --- neutron stars --- neutron star core --- gravitational waves --- stellar mergers}

%%%%%%%%%%%%%%%%%%%%%%%%%%%%%%%%%%%%%%%%%%%%%%%%%%%%%%%%%%%%%
%Introduction

\textbf{Introduction.} The nuclear equation of state (EOS) describes dense matter probed in terrestrial experiments with atomic nuclei as well as in astrophysical observations of neutron stars.
The nuclear EOS is governed by quantum chromodynamics (QCD), the theory of strong interactions, but direct calculations of dense matter in neutron stars based on QCD are not feasible at present.
Hence, the nuclear EOS has to be determined through approximate theoretical calculations or from experimental or observational data.
As a result, at densities well above nuclear saturation density, $n_{\rm sat}=0.16\fmiq$ (corresponding to a mass density of $2.7\cdot 10^{14} \textrm{g/cm}^3$), for which experimental and theoretical information are less robust, the nuclear EOS is still highly uncertain and many open questions remain, such as whether a possible phase transition to exotic phases of matter exists in nature\cite{Annala:2019puf}. 

At densities below 1-2$\ns$, the EOS and its theoretical uncertainty can be obtained from microscopic calculations based on chiral effective field theory (EFT) of QCD\cite{Hebeler:2013nza,Tews:2012fj,Lynn:2016, Drischler:2017wtt, Dris20GPB, Huth_2021}. 
To probe dense matter beyond these densities, further approaches, based on experimental and observational data, are necessary.
A very promising tool is the multi-messenger astrophysics analysis of neutron stars and their collisions, which provides access to dense neutron-rich matter not accessible in terrestrial experiments at present.
In recent years, the advent of gravitational-wave (GW) astronomy\cite{TheLIGOScientific:2017qsa} and new electromagnetic (EM) observations of neutron stars\cite{Monitor:2017mdv,Miller:2019cac,Riley:2019yda}, including the Neutron Star Interior Composition Explorer (NICER) mission of the National Aeronautics and Space Administration
(NASA)\cite{Miller:2019cac,Riley:2019yda}, led to new constraints on the EOS\cite{Bauswein:2017vtn, Annala:2017llu, Most:2018hfd, Abbott:2018exr, Radice:2018ozg, Capano:2019eae,  Dietrich:2020lps, Legred:2021hdx,Miller:2021qha,Raaijmakers:2021uju}.
However, these observations mainly probe the EOS at densities $\gtrsim 2 n_{\rm sat}$ and still carry considerable uncertainties, reflected in the ranges for predictions of neutron-star radii.
More precise or new complementary information are required to reduce the uncertainties further.

The gap between our current knowledge of the EOS stemming from nuclear theory and experiment at low densities and astrophysical observations of neutron stars at higher densities can be bridged by heavy-ion collision (HIC) experiments. 
These experiments, performed with heavy-ion beam energies of up to 2 GeV per nucleon, probe the nuclear EOS mainly in a density range of 1-2$\ns$ at present\cite{Dani02Esymm,Fevre:2015fza,Russotto:2016ucm}, representing a new source of information\cite{Tsang:2019mlz}. 

\begin{figure*}[t]
    \centering
    \includegraphics[width=\textwidth]{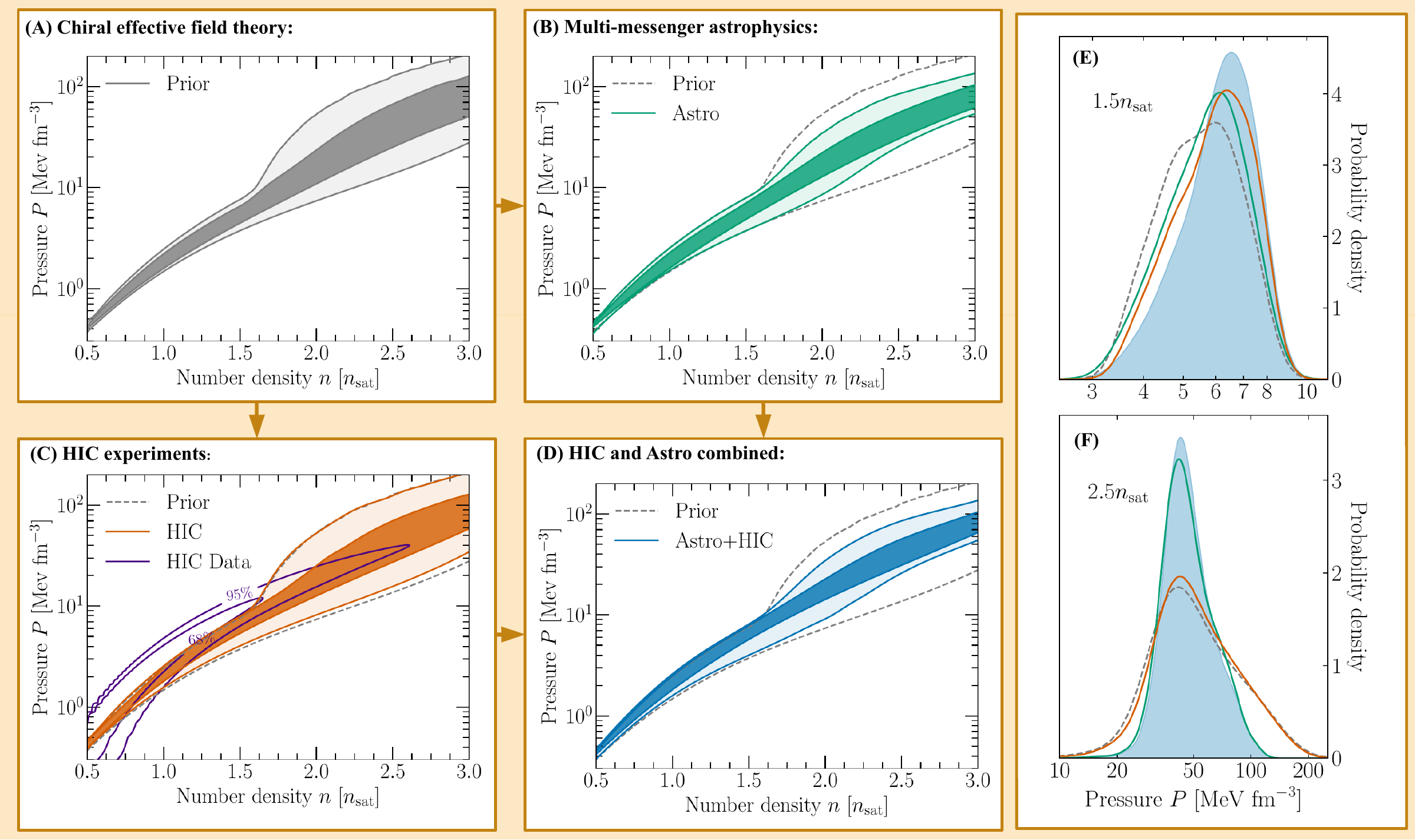}
    \caption{{\bf Constraints on the EOS of neutron-star matter.} A-D, Evolution of the pressure as a function of baryon number density for the EOS prior (A, grey), when including only data from multi-messenger neutron-star observations (B, green), when including only HIC data (C, orange), and when combining both (D, blue). 
    The shading corresponds to the 95\% and 68\% credible intervals (lightest to darkest).
    The impact of the HIC experimental constraint (HIC Data, purple lines at 95\% and 68\%) on the EOS is shown in panel C.
    In B-D, the 95\% prior bound is shown for comparison (grey dashed lines).
    E,F Posterior distributions for the pressure at $1.5 \ns$ (E) and $2.5 \ns$ (F) at different stages of our analysis (E, F), with the combined Astro+HIC region shaded in light-blue.}
    \label{fig:EOS_joint_constraint}
\end{figure*}

In this work, we perform a global analysis of the nuclear EOS including information from nuclear theory (Fig.~\ref{fig:EOS_joint_constraint}A), astrophysical observations of neutron stars (Fig.~\ref{fig:EOS_joint_constraint}B), and results from HIC experiments that were performed at the Schwerionensynchrotron 18 accelerator located at the GSI Helmholtz Centre for Heavy Ion Research\cite{Fevre:2015fza,Russotto:2016ucm} (Fig.~\ref{fig:EOS_joint_constraint}C). 
We analyse the EOS and neutron-star properties by extending our Bayesian multi-messenger astrophysics framework\cite{Dietrich:2020lps} to include information from the Four-Pi (FOPI)\cite{Fevre:2015fza} and the Asymmetric-Matter EOS (ASY-EOS) experimental campaigns\cite{Russotto:2016ucm}.
The combination of these experiments provides new constraints for neutron-rich matter in the range around 1-2$\ns$. 
We also include the EOS constraint from ref.\cite{Dani02Esymm} for symmetric nuclear matter obtained from HIC experiments at the Bevalac accelerator at Lawrence Berkeley National Laboratory and the Alternating Gradient Synchrotron at Brookhaven National Laboratory. 
In all experiments, gold nuclei were collided. 
The information from this series of HIC experiments allows us to further constrain the EOS in a density range for which theoretical calculations become less reliable.

%%%%%%%%%%%%%%%%%%%%%%%%%%%%%%%%%%%%%%%%%%%%%%%%%%%%%%%%%%%%%
%Methodology

\textbf{Nuclear theory input.}
Our analysis starts with a set of 15,000 EOSs that are constrained by nuclear theory calculations at low densities.
In particular, we utilize calculations using local chiral EFT interactions\cite{Lynn:2016, Tews:2018kmu}.
Chiral EFT is an effective theory of QCD that describes strong interactions in terms of nucleon and pion degrees of freedom using a systematic momentum expansion of nuclear forces\cite{Epelbaum:2008ga,Machleidt:2011zz}.
In particular, the EFT expansion enables estimates of theoretical uncertainties\cite{Epelbaum:2015epja,Dris20GPB}.
On the basis of local chiral two- and three-nucleon interactions, we use quantum Monte Carlo methods, which are among the most precise many-body methods to solve the nuclear many-body problem\cite{Carlson:2015}.
The breakdown scale of the chiral EFT expansion was estimated to be about 500-600~MeV/c, in which $c$ is the speed of light\cite{Dris20GPB}.
Therefore, we constrain our EOS set using chiral EFT input only up to $1.5 \ns$ (corresponding to Fermi momenta of the order of 400~MeV/c), but a variation in the range 1-2$\ns$ shows no substantial impact on our final results for neutron-star radii\cite{Essick:2020flb} (Extended Data Table~\ref{tab:prior_table}).
The 15,000 EOSs are sampled such that they span the theoretical uncertainty range of the chiral EFT calculation.

\begin{figure*}[t]
    \centering
    \includegraphics[width=\textwidth]{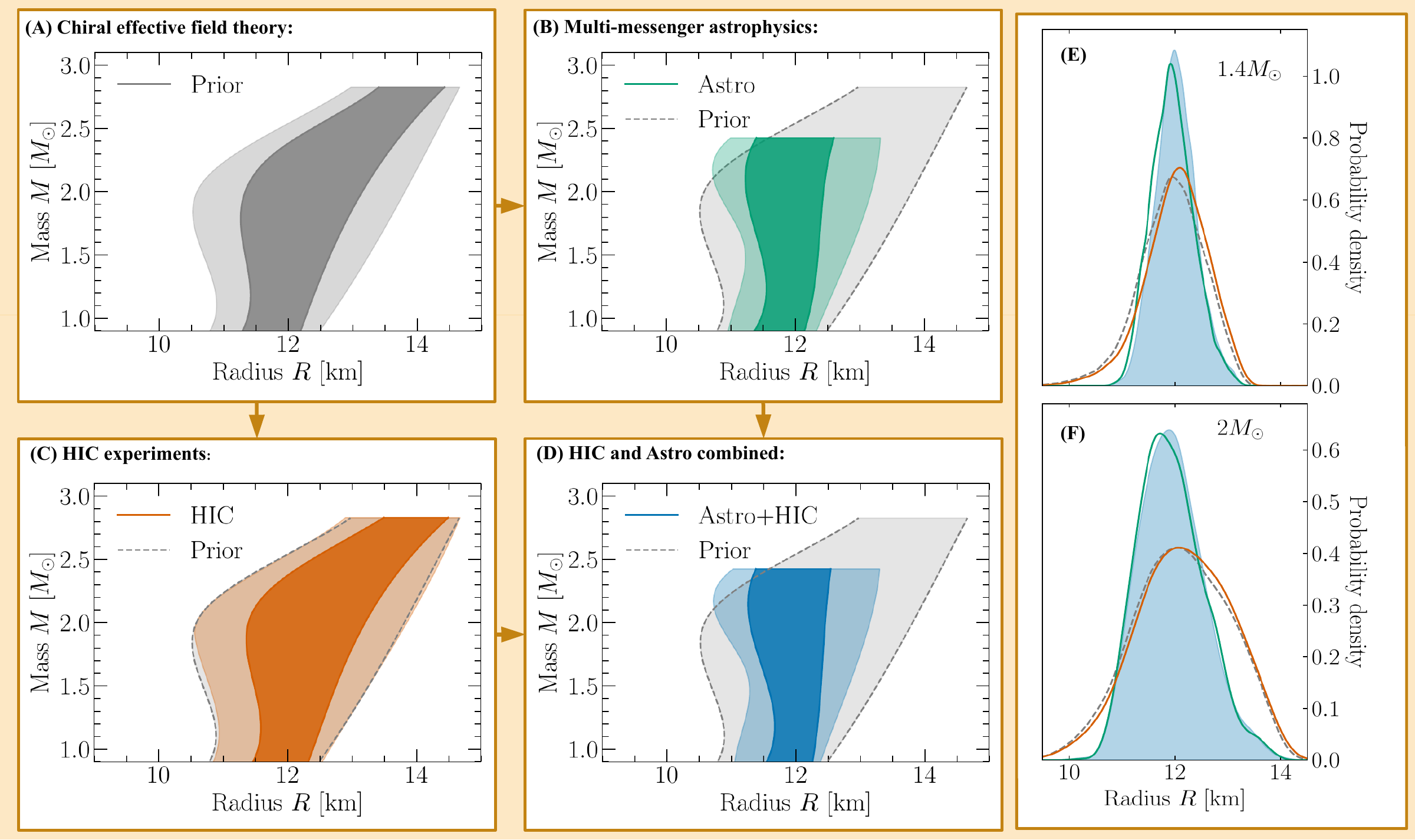}
    \caption{{\bf Constraints on the mass and radius of neutron stars.} A-D, The 95\% and 68\% credible ranges for the neutron-star radius across various masses (up to the 95\% upper bound on the maximum allowed mass, as only few EOSs support mass beyond that, which would result in an unrepresentative credible range) for the prior (A, grey), when including only multi-messenger constraints (B, green), when including only HIC experiment data (C, orange), and for the joint constraint (D, blue).
    The prior 95\% contour is shown in B-D for comparison.
    E,F, Posterior distributions for the radii of 
    $1.4M_{\odot}$ (E) and $2M_{\odot}$ (F) stars are given at different stages of our analysis, with the combined Astro+HIC region shaded in light blue.
    }
    \label{fig:MR_joint_constraint}
\end{figure*}

We extend each EOS above $1.5 \ns$ using an extrapolation in the speed of sound ($c_s$) in neutron-star matter\cite{Tews:2018chv}.
This extrapolation is constrained only by causality ($c_s\leq c$) and stability of neutron-star matter ($c_s\geq 0$).
In contrast to refs.~\cite{Annala:2017llu, Most:2018hfd}, we do not take into account any information at asymptotically high densities from perturbative QCD calculations.
In addition, at this level we require all EOSs in the prior to support neutron stars with masses of at least 1.9 solar masses ($1.9M_{\odot}$), to remove EOSs that support only neutron stars with maximum masses well below the lower limit from the combined observations of heavy pulsars\cite{Antoniadis:2013pzd,Arzoumanian:2017puf,Fonseca2021}. 
Hence, this lower bound ensures that the resulting EOS prior has reasonable support for massive-pulsar observations that we include at the first state of our Bayesian framework~\cite{Dietrich:2020lps}.
These general assumptions lead to a broad uncertainty for the EOS at higher densities (Fig.~\ref{fig:EOS_joint_constraint}A), as well as for neutron-star masses and radii (Fig.~\ref{fig:MR_joint_constraint}A).
The EOS prior is then used to analyse astrophysical observations and HIC experiments. 

\textbf{Multi-messenger astrophysics information.} 
The astrophysical data are incorporated using a Bayesian multi-messenger framework\cite{Dietrich:2020lps,Pang:2021jta}, which analyses each EOS with respect to its agreement with a variety of observational data. 
We start with the mass measurements of the massive neutron stars  PSR~J0348+0432 (ref.\cite{Antoniadis:2013pzd}) and PSR~J1614-2230 (ref.\cite{Arzoumanian:2017puf}), to obtain a lower bound on the maximum mass, and the constraint on the maximum mass of neutron stars derived from the binary neutron-star collision GW170817 (refs.~\cite{Margalit:2017dij,Rezzolla:2017aly}) in which a black hole was formed after the coalescence, to obtain an upper bound on the maximum mass. 
Information obtained from X-ray pulse-profile modelling of PSR~J0030+0451 and PSR~J0740+6620 using data from NICER and the X-ray Multi-Mirror Mission (XMM-Newton)\cite{Miller:2019cac,Miller:2021qha,Riley:2021pdl} are incorporated. 
Moreover, we use Bayesian inference techniques to analyse GW information from the two neutron-star mergers GW170817 (ref.\cite{TheLIGOScientific:2017qsa}) and GW190425 (ref.\cite{Abbott:2020uma}) by matching the observed GW data with theoretical GW models that depend on neutron-star properties.
For our analysis, we use a GW model\cite{Dietrich:2019kaq} that is an improved version of the main waveform model used by the Laser Interferometer Gravitational-Wave
Observatory/Virgo Collaboration for the study of GW170817 (ref.\cite{Abbott:2018wiz}) and GW190425 (ref.\cite{Abbott:2020uma}).
Similarly to the GW analysis, we also include information from the kilonova AT2017gfo (ref.\cite{Monitor:2017mdv}) associated with the GW signal.
Kilonovae originate from the radioactive decay of heavy atomic nuclei created in nucleosynthesis processes during and after the merger of neutron stars, and are visible in the optical, infrared and ultraviolet spectra.
The electromagnetic observations are analysed with full radiative transfer simulations\cite{Bulla:2019muo} to extract information from the observed light curve and spectra\cite{Coughlin:2017ydf}.

The above astrophysical information leads to important constraints on the neutron-star EOS, as shown in Fig.~\ref{fig:EOS_joint_constraint}B.
The constraints are strongest above $1.5 \ns$, for which the extrapolation in the speed of sound is used for the EOSs. 
The high-density astrophysical constraints affect mostly the high-mass region in the mass--radius plane and exclude the stiffest EOSs that lead to the largest radii. (Fig.~\ref{fig:MR_joint_constraint}B).

\textbf{Data from HIC experiments.}
To further constrain the EOS, we implement data from HIC experiments. 
The FOPI\cite{Fevre:2015fza} and ASY-EOS\cite{Russotto:2016ucm} experiments performed at GSI provide information respectively on the symmetric nuclear matter EOS (that is,  matter with the same amount of protons and neutrons), and on the symmetry energy, which describes the energy cost of changing protons into neutrons in nuclear matter.
For both experiments, $^{197}$Au nuclei were collided at relativistic energies (0.4 to 1.5 GeV per nucleon), forming an expanding fireball in the collision region.
This expansion is dictated by the achieved compression and therefore depends on the EOS of hot and dense matter. 
Owing to the initial neutron-to-proton asymmetry of the Au--Au system, the expansion of the emitted nucleons is sensitive to the nuclear symmetry energy. 
Constraints on the symmetry energy (from ASY-EOS) can be translated into a constraint on the pressure of neutron-star matter as a function of the baryon density when empirical information on symmetric nuclear matter from experiments (FOPI) with atomic nuclei is used. 
In addition to the GSI experiments, we include constraints on the pressure of symmetric nuclear matter at larger densities obtained from model calculations of ref.\cite{Dani02Esymm} that were used to analyse experimental data from Lawrence Berkeley National
Laboratory and Brookhaven National Laboratory in which $^{197}$Au nuclei were collided at energies up to 10 GeV per nucleon.
These are sensitive to higher densities, 2-4.5$\ns$, but we include their constraints only up to $3\ns$, where the sensitivity of the ASY-EOS experiment ends. 
We find that the inclusion of this further constraint has only minimal impact, but keep it to ensure the completeness of our study (Supplementary Information).

In Fig.~\ref{fig:EOS_joint_constraint}C, we show the combined HIC experimental constraints (labelled HIC Data) at $68\%$ and $95\%$ credibility as well as the resulting posterior distribution for the neutron-star EOS. 
Whereas the FOPI experiment delivers an EOS constraint for symmetric nuclear matter at densities in the range 1-3$\ns$, the ASY-EOS experiment probes the symmetry energy roughly between 1 and 2$\ns$. 
The HIC pressure-density constraint includes various sources of uncertainties.
First, it includes systematic and statistical uncertainties of the experiments and the analysis of its data\cite{Fevre:2015fza, Russotto:2016ucm}.
We have explicitly checked the robustness of our results when varying the details of the analysis and models used, and generally found that our results do not substantially depend on individual model choices (Extended Data Tabs.~\ref{tab:cse_pwp_comparison}, Extended Data Fig.~\ref{fig:sensitivity_curves} and Supplementary Information).
Second, when extracting the HIC constraint on neutron-star matter, we vary nuclear matter properties, such as the incompressibility parameter and the symmetry energy at $\ns$, according to the measurements from FOPI and ASY-EOS. 
We have explicitly checked that increasing these uncertainties in agreement with theoretical estimates\cite{Huth_2021} leads to only minor changes of our final results (Extended Data Tab.~\ref{tab:inflated_errors}). 

To enforce the ASY-EOS constraints only at densities for which the experiment is sensitive, we use the sensitivity curve for neutrons and charged particles\cite{Russotto:2016ucm} as a prior for the probed density range.
We have checked the variation of our results for alternative choices of the sensitivity curve\cite{Russotto:2016ucm} and found that this has no substantial impact on our final results (Extended Data Tab.~\ref{tab:sensitivity_table}).
We find that the HIC constraints tend to prefer EOSs stiffer than the ones favoured by astrophysical observations (that is, EOSs that have higher pressures at densities up to 2$\ns$; Fig.~\ref{fig:EOS_joint_constraint}C, E).

We note that results of the ASY-EOS experiment, in their sub-saturation density extension, are compatible with recent experimental findings from isobaric analog states supplemented with further constraints from neutron-skin data\cite{Dani14IAS_esym}, HICs using isospin-diffusion observables measured in mid-peripheral collisions of Sn isotopes\cite{Tsan09HIC_esym}, and other nuclear structure information\cite{Zhang:2013wna,Brown:2013prl}.
More recently, the S$\pi$rit campaign at RIKEN has identified spectral yield ratios of charged pions in collisions of various tin isotopes near threshold as sensitive probes of the slope of the symmetry energy near and beyond nuclear saturation density\cite{Estee:2021khi}. 
The obtained value is compatible with the ASY-EOS result but offers no further strong constraint at present owing to its large uncertainty\cite{Estee:2021khi,yong2021symmetry}.

\textbf{Combining micro- and macroscopic collisions.} 
The final EOS constraints are obtained through the combination of both the HIC information and astrophysical multi-messenger observations (Fig.~\ref{fig:EOS_joint_constraint}D). 
The multi-messenger data rule out the most extreme EOS behavior, and the HIC data favour larger pressures around 1-1.5$\ns$, for which the experimental sensitivity is highest. This is similar to the effect of recent NICER observations on the EOS\cite{Miller:2021qha,Raaijmakers:2021uju}.
Hence, the two complementary approaches, HIC experiments and astrophysical observations show a remarkable agreement (Fig.~\ref{fig:EOS_joint_constraint}E). 
At low densities, HIC results have a clear impact on the total posterior for the EOS, whereas the EOS at higher densities ($\gtrsim 2 \ns$) is mostly determined by astrophysical observations.
At these densities, HIC results deviate only mildly from the prior (Fig.~\ref{fig:EOS_joint_constraint}F). 
This is also reflected in the radii of neutron stars shown in Fig.~\ref{fig:MR_joint_constraint}E, F.
As astrophysical observations mainly probe neutron stars with $M\gtrsim 1.4M_\odot$, for which the probed densities are higher, HIC information influences the radii of these neutron stars to a smaller degree. 
The radius of low-mass stars with $M\approx 1.0M_\odot$, on the other hand, is also constrained by HIC information.
Our final result for a typical 1.4$M_\odot$ neutron star is $12.01^{+0.37}_{-0.38} \rm{km}$ at $68\%$ uncertainty ($12.01^{+0.78}_{-0.77} \rm{km}$ at $95\%$ uncertainty; Tab.~\ref{tab:main_results}).
Comparing this value to the result without any HIC information, $11.93^{+0.39}_{-0.41} \rm{km}$ at $68\%$ confidence, highlights the benefit of combining these various sources of information in a statistically robust framework. We find that the HIC information has a high impact on the EOS at densities below $1.5\ns$ (Supplementary information).
Finally, we quantify the possibility for the presence of a strong first-order phase transition to a new phase of QCD matter in the core of neutron stars.
For this, we calculate the Bayes factor in favour of the presence of such a phase transition against its absence, and find it to be $0.419 \pm 0.012 < 1$. 
Therefore, its presence is slightly disfavoured given current astrophysical and experimental data.

\begin{table}[]
    \begin{center}
    \renewcommand{\arraystretch}{1.3}
    \caption{{Final constraints on the pressure and the radius of neutron stars.}}
    \label{tab:main_results}
    \begin{tabular}{ccccc}
        \hline
         & Prior & Astro only & HIC only & Astro + HIC \\
         \hline
        $P_{1.5\ns}$ & $5.59^{+2.04}_{-1.97}$ & $5.84^{+1.95}_{-2.26}$ & $6.06^{+1.85}_{-2.04}$ & $6.25^{+1.90}_{-2.26}$\\
        $R_{1.4}$ & $11.96^{+1.18}_{-1.15}$ & $11.93^{+0.80}_{-0.75}$ & $12.06^{+1.13}_{-1.18}$ & $12.01^{+0.78}_{-0.77}$\\
        \hline
    \end{tabular}
    \end{center}
    Comparison of the pressure in $\MeV\fmiq$ at $1.5\ns$ and the radius in km of a $1.4M_\odot$ neutron star (median with the 95\% credible interval) when including only astrophysical constraints, only HIC experimental data, and the combination of both.
\end{table}

To summarize, the interdisciplinary analysis of EOS constraints from HIC experiments and multi-messenger astrophysics shows remarkable agreement between the two, and provides important information to constrain the nuclear EOS at supra-saturation densities. 
Going forward, it is important that both statistic and systematic sources of uncertainty for HIC experiments are further improved.
For example, the impact of choosing different quantum molecular dynamics models when analysing HIC experiments needs to be further investigated (Extended Data Fig.~\ref{fig:S0_gammaasy} and \ref{fig:sensitivity_curves}), and advancing HIC experiments to probe higher densities, above 2-3$\ns$, will be key (Extended Data Tab.~\ref{tab:expolratory_table}).
Combining the latter with a reduction of experimental uncertainties, data from HICs have great potential to provide complimentary EOS information, bridging nuclear theory and astrophysical observations.
In the next few years, the ASY-EOS-II and Compressed Baryonic Matter experiments at the upcoming Facility for Antiproton and Ion Research at GSI will provide a unique opportunity to study nuclear matter at densities probed in the core of neutron stars and their mergers, and might detect new phases of QCD matter, possibly involving hyperons and, ultimately, the transition to a deconfined quark matter phase at the highest densities (see, for example, refs.~\cite{Orsaria:2019ftf,Brandes:2021pti}).
Together with experiments at the Rare Isotope Beam Facility at RIKEN in Japan and the Nuclotron-based Ion Collider Facility in Russia, the robust combination of experimental HIC constraints and astrophysical observations has the potential to revolutionize our understanding of the EOS. 

%\bibliographymain{refs}
%\bibliographystylemain{naturemag}

%\FloatBarrier
%\clearpage

%%%%%%%%%%%%%%%%%%%%%%%%%%%%%%%%%%%%%%%%%%%%%%%%%%%%%%%%%%%%%

\section*{\large Methods}

\noindent
\textbf{Nuclear EOSs from chiral EFT}

The EOS set used in this work is constrained at low densities by microscopic calculations of neutron matter using interactions from chiral EFT. 
In these microscopic calculations, the Schr\"odinger equation for the many-body system is solved numerically. 
This requires a nuclear Hamiltonian and a method to solve the Schr\"odinger equation with controlled approximations.

To obtain the Hamiltonian describing the dense matter EOS studied in this work, we use chiral EFT.
Chiral EFT is a low-energy effective theory of QCD, and describes strong interactions in terms of nucleon and pion degrees of freedom instead of quarks and gluons\cite{Epelbaum:2008ga,Machleidt:2011zz}.
To construct the interactions, the most general Lagrangian in terms of nucleons and pions, consistent with all symmetries of QCD, is expanded in powers of momenta.
Using a power counting scheme, the individual contributions are arranged according to their importance.
By going to higher orders, the description of interactions becomes more precise, but the individual contributions become more involved.
The chiral EFT Lagrangian explicitly includes pion-exchange interactions among nucleons whereas all high-energy details that are not explicitly resolved are expanded in terms of general contact interactions.
These are accompanied by low-energy couplings, which are fitted to experimental data.

Chiral EFT interactions have several benefits over phenomenological interaction models: they naturally include many-body forces consistent with two-nucleon interactions, they can be systematically improved, and they enable theoretical uncertainty estimates\cite{Epelbaum:2015epja,Dris20GPB}.
The latter of these can be extracted from order-by-order calculations and are important when analysing astrophysical observations for which interactions are extrapolated to conditions that cannot be recreated in experiments at present. 

In this work, we constrain our EOSs with theoretical calculations at zero temperature using local chiral EFT interactions\cite{ Lynn:2016,Gezerlis:2013,Gezerlis:2014,Tews:2015ufa}.
We use quantum Monte Carlo methods\cite{Carlson:2015}, in particular the auxiliary-field diffusion Monte Carlo method, which are among the most precise many-body methods to solve the nuclear many-body problem.
The results of these calculations provide constraints on the EOS up to densities of around $2\ns$ (ref.\cite{Tews:2018kmu}).

The region of applicability of the chiral EFT expansion is determined by the so-called breakdown scale, which is estimated to be of the order of 500-600~MeV/c (ref.\cite{Drischler:2020yad}).
Hence, the chiral EFT expansion breaks down at densities $\gtrsim 2 \ns$, indicated by increasing uncertainty estimates between 1 and 2$\ns$.
At these densities, high-energy physics that is encoded in short-range contact interactions needs to be explicitly taken into account.
Therefore, chiral EFT cannot be used to constrain the EOS at higher densities as probed in the cores of neutron stars. 
To extend the EOS to these densities, we use a general extrapolation scheme in terms of the speed of sound\cite{Tews:2018chv} (see also ref.\cite{Greif:2018njt}).

To construct the neutron-star EOS set, we first extend our chiral EFT calculation to $\beta$-equilibrium and add a crust\cite{Tews:2016ofv}.
We use microscopic input up to $1.5 \ns$ to constrain the EOS, but a variation in the range 1-2$\ns$ shows no substantial impact on our final results for neutron-star radii\cite{Essick:2020flb}.
Above this density, we sample a set of six randomly distributed points in the speed of sound plane at baryon densities between 1.5$\ns$ and 12$\ns$, enforcing $0\leq c_s\leq c$ at each point. 
A variation of the number of sampled points between 5 and 10 does not impact our findings.
We then connect these points by line segments, reconstruct the EOS, and solve the Tolman-Oppenheimer-Volkoff equations to extract neutron-star properties.
For each EOS, we also construct a partner EOS that includes a segment with vanishing speed of sound to explicitly simulate strong first-order phase transitions.
We sample the onset density and width of this segment randomly. 

Our EOS set includes 15,000 different EOS samples for which the prior on the radii of neutron stars is naturally determined by the EOS expansion scheme.
We have explicitly checked the differences among a prior uniform in the radius of a typical 1.4$M_{\odot}$ neutron star and the `natural' prior and found only minor changes once astrophysical and HIC data are included (Extended Data Tab.~\ref{tab:prior_table}).

Recently, first results for the EOS of symmetric nuclear matter between 3 and 10$\ns$ from functional renormalization group calculations that are based on QCD became available\cite{Leon19fRGeos}.
This offers a very promising future tool to constrain dense neutron-star matter when calculations for asymmetric matter will become available. 

\noindent
\textbf{Multi-messenger analysis of astrophysical data}

To constrain the set of EOSs derived from chiral EFT with astrophysical data, we use a multi-step procedure in which results from individual steps are used as a prior for the next part of the analysis\cite{Dietrich:2020lps} (Extended Data Fig.~\ref{fig:astro_steps}).
First, we incorporate constraints on the maximum mass of neutron stars.
For this, we implement the mass measurements of the heavy radio pulsars PSR~J0348+0432 (ref.\cite{Antoniadis:2013pzd}) and  
PSR~J1614-2230 (ref.\cite{Arzoumanian:2017puf}). As we make use of the NICER and XMM mass-radius information of PSR~J0740+6620 (refs.\cite{Miller:2021qha,Riley:2021pdl}) at a later stage, we do not include the mass measurement of PSR~J0740+6620 (ref.\cite{Fonseca2021}) to avoid double counting. 
The combination of these observations\cite{Dietrich:2020lps,Tews:2020ylw} of high-mass neutron stars provides a lower bound on the maximum mass of neutron stars.
By contrast, an upper bound of the maximum mass is obtained from the observation of the merger remnant of the neutron-star merger GW170817 (ref.\cite{Rezzolla:2017aly}). 
Among other arguments, the observation of a bright, red kilonova component and the observation of a short gamma-ray burst 2 s after the merger of the two neutron stars indicate that the remnant experienced a delayed ($\mathcal{O}(100\rm ms)$) collapse to a black hole, so that an upper limit on the maximum mass can be derived. 
The combined estimate of the maximum mass, $2.21^{+0.10}_{-0.13}M_{\odot}$ at $68\%$ uncertainty, already provides important information about the internal structure of neutron stars and disfavours both too stiff and too soft EOSs (that is, EOSs with too large and too small pressures, respectively).

In the next step, we incorporate NICER's mass and radius measurement of PSR~J0030+0451 (ref.\cite{Miller:2019cac}) and PSR~J0740+6620 (ref.\cite{Miller:2021qha,Riley:2021pdl}). 
NICER, located on board of the International Space Station, is a NASA telescope measuring the X-ray pulse profile of pulsars. 
By correlation of the observed profile and brightness with theoretical predictions, it is possible to extract information on the configuration (for example, on the location and properties of hotspots on the neutron-star surface, the rotation rate of the star, and its compactness, which determines the light bending around the pulsar).
This information enables constraints on the pulsar's mass and radius. 
In addition to NICER, the XMM-Newton telescope\cite{Struder:2001bh, Turner:2000jy} has been used for the analysis of PSR~J0740+6620 (ref.\cite{Miller:2021qha}) to improve the total flux measurement.
For PSR~J0740+6620, we average over the results obtained in refs.\cite{Miller:2021qha,Riley:2021pdl}, whereas for PSR~J0030+0451 we use only results of ref.\cite{Miller:2019cac}.

Next, we analyse the GW signal emitted from the binary neutron-star merger GW170817 (ref.\cite{TheLIGOScientific:2017qsa}), as well as its observed kilonova AT2017gfo (ref.\cite{Monitor:2017mdv}). 
Finally, we also incorporate the second confirmed GW signal from a binary neutron-star merger GW190425 (ref.\cite{Abbott:2020uma}). For GW170817 and GW190425, we assumed both of them to be emitted by binary neutron star mergers.
To test the robustness of the GW analysis, we have explored a number of different GW models and found only a minimal impact on the final EOS constraint\cite{Dietrich:2020lps}. 
Results shown in the main text are obtained using the parallel bilby software\cite{Smith:2019ucc}
and the waveform model IMRPhenomPv2\_NRTidalv2 (ref.\cite{Dietrich:2019kaq}) for cross-correlation with the observed data\cite{TheLIGOScientific:2017qsa}. 
IMRPhenomPv2\_NRTidalv2 is an updated model of the waveform model used in previous analyses by the the Laser Interferometer
Gravitational-Wave Observatory (LIGO)/Virgo Collaboration\cite{Abbott:2020uma,Abbott:2018wiz} and, hence, allows for a more accurate measurement of tidal effects. 
The likelihood function for the GW analysis $\mathcal{L}_{\rm GW}$ is given by\cite{Veitch:2014wba}
\begin{equation}
    \mathcal{L}_{\rm GW} \propto \exp \left(-2 \int df \frac{|\tilde{d}(f) - \tilde{h}(f)|^2}{S_n(f)}\right),
\end{equation}
in which $\tilde{d}(f)$, $\tilde{h}(f)$, and $S_n(f)$ are the observed data, the waveform template, and the power spectral density of the noise, respectively. 
To ensure full coverage of the binary neutron stars' inspiral signal, we have analysed the data up to 2,048Hz. 
To avoid the low frequency noise-wall in the detectors, a low frequency bound of 20 Hz is used. 

Similarly, we use Bayesian inference to analyse the observed kilonova AT2017gfo. The likelihood function for the light curve analysis $\mathcal{L}_{\rm EM}$ is given by\cite{Coughlin:2018miv}
\begin{equation}
    \mathcal{L}_{\rm EM} \propto \chi^2_1 \left(\sum_{ij}\frac{1}{n_j - 1}\left(\frac{m^{j}_{i} - m^{j, \rm{est}}_{i}}{\sigma^{j}_{i}}\right)^2\right),
\end{equation}
in which $m^{j, \rm{est}}_{i}$ are the estimated or theoretically predicted apparent magnitudes for a given filter $j$ (a passband for a particular wavelength interval) at observation time $t_i$ with $n_j$ data points for filter $j$. Moreover, $m^{j}_{i}$ and $\sigma^{j}_{i}$ are the observed apparent magnitude and its corresponding statistical uncertainties, respectively. For this analysis, the probability distribution of a chi-squared distribution with a degree of freedom of 1, $\chi^2_1$, is taken as the likelihood measurement. 
To reduce the systematic error of the kilonova modelling below the statistical error, a further uncertainty of 1 mag is added to the measurement error.
To analyse AT2017gfo, we use the radiative transfer code POSSIS\cite{Bulla:2019muo} to produce grids of light curves for multidimensional kilonova models with the following free parameters: the dynamical ejecta mass, the disk wind ejecta mass, the opening angle of the lanthanide-rich dynamical-ejecta component, and the viewing angle. To enable inference, we combine the grid with a framework combining Gaussian process regression and singular value decomposition\cite{Coughlin:2018fis} to compute generic light curves for these parameters.
To connect the ejecta parameters, which determine the exact properties of the light curve, with the binary neutron-star system parameters, we assume that the total ejecta mass is a sum of two components: dynamical ejecta, released during the merger process through torque and shocks, and disk-wind ejecta. 
Both components, the dynamical ejecta\cite{Coughlin:2018fis} and the disk-wind ejecta\cite{Dietrich:2020lps}, are correlated to source parameters of the binary neutron-star system based on numerical relativity simulations\cite{Dietrich:2020lps,Dietrich:2016fpt,Coughlin:2018fis}. 

\noindent
\textbf{Constraining the symmetric nuclear matter EOS at high density with HICs}

Over the last two decades, major experimental efforts have been devoted to measuring the nuclear EOS with HIC experiments performed at relativistic incident energies\cite{Dani02Esymm,Fuchs:2005yn,Zhang:2020dvn}.
These collisions of atomic nuclei form a hot, dense fireball of hadronic matter in the overlapping region, which expands in time and reaches the surrounding detectors as baryons and mesons. 
The phase-space distribution of particles flowing from the fireball during the expansion phase is strongly dictated by the compression achieved in the colliding region and is, therefore, sensitive to the EOS of the hot and dense nuclear matter created in the collision. 
Important progress has been made recently in modelling intermediate-energy HICs but theoretical uncertainties still remain\cite{An:2021wof,Colonna:2021xuh}.
In the present analysis, results obtained with different models are found to be compatible within their quoted errors.

The so-called elliptic flow ($v_2$) of emerging particles is the main observable, which has been used to experimentally constrain symmetric nuclear matter at supranuclear densities with HICs. 
It is described by the second moment of the Fourier expansion of the distribution of the azimuthal angle $\Phi$ of the emitted particles with respect to that of the reaction plane $\Phi_{\rm RP}$
\begin{equation}
\begin{aligned}
    \frac{d\sigma(y,p_{t})}{d\Phi}= &C(1+2v_{1}(y,p_t)\cos(\Phi-\Phi_{\rm RP})\\
    &+2v_{2}(y,p_t)\cos~2(\Phi-\Phi_{\rm RP})+...)\,,
\end{aligned}
\end{equation}
in which all expansion coefficients $v_{n}$ are functions of longitudinal rapidity $y = \frac{1}{2}\ln\left(\frac{E+p_z}{E-p_z}\right)$, with $p_z$ being the momentum along the beam axis and $E$ being the total energy, and of transverse momentum $p_{t}=\sqrt{p_x^2+p_y^2}$ of the particle, with $p_x$ and $p_y$ denoting the momentum components perpendicular to the beam axis. 

In the experiment, the orientation of the reaction plane is event-wise reconstructed from the azimuthal distribution of particles recorded in the forward and backward hemispheres, and the Fourier coefficients are corrected for the finite resolution of this procedure\cite{Andronic:2006ra}.
The coincident particle and fragment emissions are also used for the reconstruction of the impact parameter of each reaction event\cite{Russotto:2016ucm}.
A positive elliptic flow $v_2$ indicates a preferred emission in the reaction plane whereas a negative flow indicates an emission out of the reaction plane.

It has been shown that the elliptic flow $v_{2}$ of protons emitted at rapidities intermediate between projectile and target rapidity (mid-rapidity) in HICs at incident energies of several hundred MeV per nucleon offers the strongest sensitivity to the nuclear EOS\cite{Fevre:2015fza,Dani02Esymm,FOPI:2011aa}, as evident from calculations made with various transport models. 
This dependence on the nuclear EOS is predicted by quantum molecular dynamics (QMD)\cite{Fevre:2015fza,Hartnack:1997ez, FOPI:2011aa, Wang:2018hsw} and Boltzmann-Uehling-Uhlenbeck\cite{Dani02Esymm} models. The origin of the phenomenon has been investigated in detail elsewhere\cite{LeFevre:2016vpp}.
As shown in ref.\cite{Dani02Esymm}, at higher beam energies between 1 and 10 GeV per nucleon, the sensitivity of the directed flow $v_1$ to the stiffness of the EOS of symmetric nuclear matter becomes comparable to that of $v_2$.
Overall, from HICs performed at incident beam energies of a few hundred MeV per nucleon up to around 10 GeV per nucleon, the flow data indicate an EOS for symmetric nuclear matter with an incompressibility $K$ below 260~MeV. 
Using FOPI data on the elliptic flow in gold-gold collisions between 400~MeV and 1.5~GeV per nucleon, thanks to the broad acceptance of the detector, an enhanced precision in the determination of the EOS could be achieved. Including the full rapidity and transverse momentum dependence of the elliptic flow of protons and heavier isotopes\cite{Fevre:2015fza} in the analysis with the Isospin-QMD (IQMD) transport model, the incompressibility was determined as $K=190\pm30$~MeV. 
This result was confirmed by interpreting the same data with three Skyrme energy-density functionals introduced into the ultrarelativistic QMD (UrQMD) transport model\cite{Wang:2018hsw}, leading to $K=220\pm40$~MeV. 
The interval of confidence used in the present study, $K=200\pm25$~MeV, reflects both predictions. 
The densities probed were estimated to range between 1 and 3$\ns$ by analysing the densities effective in building the elliptic flow in IQMD simulations\cite{Fevre:2015fza}. 
Note that the constraints deduced from the analysis of elliptic flow are compatible with earlier findings of the Kaon Spectrometer Collaboration obtained from comparisons of QMD predictions with experimental $K^+$ meson production yields from gold-gold and carbon-carbon collisions performed at GSI between 0.6 and 1.5 GeV per nucleon\cite{Sturm:2000dm,Fuchs:2001gv}.

\noindent
\textbf{The ASY-EOS experiment to measure the symmetry energy}
 
Nuclear experiments can help to constrain the EOS of neutron matter (see, for example, the PREX experiment measuring the neutron-skin thickness in lead nuclei~\cite{PREXII, Reed:2021nqk, Essick:2021kjb, Biswas:2021yge}).
It has been suggested\cite{Li:2002qx} that flows of particles in HICs can be used to constrain the EOS of neutron matter through the symmetry energy at supra-saturation density.
However, nuclear matter that can be studied in laboratory experiments using HICs is not very neutron-rich. 
Therefore, the effect of the symmetry energy on $v_{2}$ remains small, close to or below the uncertainties of the main contribution of the symmetric nuclear matter EOS. 
To enhance observable effects related to the symmetry energy, the use of the 
elliptic flow ratio of particles with large isospin difference, ideally the ratio for neutrons over protons $v_2^{\rm np} = v_2^{\rm n} / v_2^{\rm p}$, was proposed\cite{Russotto:2011hq}. 
This method has been adopted for the ASY-EOS experiment performed at GSI in Darmstadt, studying collisions of gold nuclei of 400~MeV per nucleon incident energy and gold targets. The description of the experiment and the analysis with the UrQMD transport model are given in detail elsewhere\cite{Russotto:2016ucm}
ASY-EOS benefited from the Large-Area Neutron Detector (LAND)\cite{Blaich:1991LAND} permitting the detection of neutrons and charged particles within the same acceptance. LAND was placed to cover mid-rapidity emissions over a large $p_t$ interval.  
Its isotopic resolution in this experiment was not sufficient to uniquely identify protons. Elliptic flow ratios as a function of $p_t$ were, therefore, determined for neutrons with respect to all charged particles within the LAND acceptance.
We note that for the selected collisions (central up to semi-central) and angular region, the yield of charged particles consists of light isotopes, mainly protons (around 50\%) according to FOPI data for the same reaction.
Confronted with UrQMD transport model predictions (and confirmed with other models, IQMD\cite{Hartnack:1997ez} and Tübingen QMD\cite{Cozma:2017bre}), the resulting flow ratio enabled deduction of a constraint for the symmetry energy, which is so far the most precise for supra-saturation densities obtained from HICs. The results are detailed in the following section. 
As indicated by QMD model predictions, densities probed by the elliptic flow ratio in the ASY-EOS experiment extend up to about $2 \ns$.

\noindent
\textbf{Implementation of nuclear EOS constraints from HICs}

For analysing the experimental elliptic flow data, an EOS functional needs to be fed into the QMD simulations for both symmetric and asymmetric nuclear matter.
This is given by the parameterization for the energy per particle
\begin{equation}
\dfrac{E}{A}(n,\delta)\approx\dfrac{E}{A}(n,0)+S(n)\delta^2 \,,
\label{eq:expandE}
\end{equation}
with the baryon density $n=n\n+n\p$ and the isospin asymmetry $\delta=(n\n-n\p)/n=1-2x$, in which $n\n$ and $n\p$ are the neutron and proton densities, respectively, and $x=n\p/n$ is the proton fraction. 
$E/A(n,0)$ is the energy of symmetric nuclear matter, $E/A(n,1)$ corresponds to pure neutron matter, and $S(n)$ is the symmetry energy defined here as the difference between the two. 
For the analysis of the FOPI experiment, the first term in Eq.~\eqref{eq:expandE} has been parameterized with
\begin{equation}
    \dfrac{E}{A}(n,0)=\dfrac{3}{5}\left(\dfrac{n}{\ns}\right)^{2/3}E_F + \dfrac{\alpha n}{2\ns}+\dfrac{\beta}{\gamma +1}\left(\dfrac{n}{\ns}\right)^{\gamma}\,,
    \label{eq:asy_ESNM}
\end{equation}
with the saturation density $\ns$, the Fermi energy $E_F$, and in which the parameters $\alpha,\beta,$ and $\gamma$ are fixed by the incompressibility $K$, the binding energy $B$ of symmetric nuclear matter at $\ns$, and the condition that the pressure of symmetric nuclear matter is zero at saturation density:
\begin{align}
    \alpha &= -2\left(\dfrac{K+\frac{6E_F}{5}}{9\left(\gamma-1\right)}+\dfrac{2}{5}E_F \right)\,,\nonumber\\
    \beta &= \left( K+\dfrac{6}{5}E_F \right)\dfrac{\gamma + 1}{9\gamma\left(\gamma-1\right)}\,,\quad
        \label{eq:Skyrme_paras}
    \gamma = \dfrac{K + \frac{6E_F}{5}}{9\left(\frac{E_F}{5}+B\right)}\,.
\end{align}

In the ASY-EOS analysis, the $S(n)$ term of Eq.~\eqref{eq:expandE} has been parameterized as
\begin{equation}
    S(n) = E_\mathrm{kin,0}\left(\dfrac{n}{\ns}\right)^{2/3} + E_\mathrm{pot,0}\left(\dfrac{n}{\ns}\right)^{\gasy}\,.
    \label{eq:Esym_QMD}
\end{equation}
At saturation density, the kinetic part has been set to $E_\mathrm{kin,0}=12\MeV$ and $E_\mathrm{pot,0}=\esym-E_\mathrm{kin,0}$.
The parameter $\gasy$ was extracted from fits to experimental data of the $p_t$ dependence of the elliptic flow ratio of neutrons over charged particles around mid-rapidity.
In particular, this results in $\gasy=0.68\pm 0.19$ for $\esym=31$~MeV and $\gasy=0.72\pm 0.19$ for $\esym=34$~MeV (Extended Data Fig.~\ref{fig:PNM_constraints} for a comparison with microscopic neutron matter calculations).
Here we interpolate $\gasy$ assuming a linear function with $\esym$, for which the uncertainty is chosen to be $0.19$ independent of $\esym$.
We have studied the behavior of $\gasy$ as a function of $S_0$ for two different QMD models (Extended Data Fig.~\ref{fig:S0_gammaasy}) and confirmed that the linear interpolation in the $\esym$ range is suitable. 

The pressure constraint is given by the density derivative of the energy per particle of equation~\eqref{eq:expandE},
\begin{equation}
    P(n, \delta) =n^2\dfrac{\partial E/A(n, \delta)}{\partial n}\,,
    \label{eq:pressure}
\end{equation}
and depends on $n$, $\delta$, $\ns$, $B$, $K$, and $\esym$.
We enforce this constraint only at densities for which the experiment is sensitive.
The density region of the HIC constraint is set by the sensitivity of the neutrons-over-charged-particles flow ratio determined for the ASY-EOS experiment\cite{Russotto:2016ucm} (see also the previous section). 
This sensitivity curve covers the density range from $0.5\ns$ up to $3\ns$ and peaks between $\ns$ and about 2$\ns$, for which the experiment is most sensitive.

Neutron-star matter is composed of neutrons, protons, electrons and muons in $\beta$-equilibrium. 
To apply the ASY-EOS constraint to neutron stars, we need to determine the proton fraction $x_\mathrm{ASY\mbox{-}EOS}$ accordingly. 
For simplicity, we neglect muons because they only have a small impact on the neutron-star EOS in the considered density range. 
Then, the density of electrons is equal to the proton density owing to local charge neutrality, and the proton fraction $x$ at a given baryon density $n$ is fixed by the $\beta$-equilibrium condition,
\begin{align}
\mu\n(n,x) &=\mu\p(n,x)+\mu_\mathrm{e}(n_\mathrm{e}=xn)\,,
\end{align}
in which $\mu_{\text{n},\text{p},\text{e}}$ is the chemical potential of the respective particle species.
We calculate the neutron and proton chemical potentials consistently with equations~\eqref{eq:expandE}-\eqref{eq:Esym_QMD}.
Electrons are modelled as an ultrarelativistic degenerate Fermi gas with pressure $P_\mathrm{e}=E_\mathrm{e}/(3V)$, energy density $E_\mathrm{e}/V=\hbar c(3\pi^2 n_\mathrm{e})^{4/3}/(4\pi^2)$, and chemical potential $\mu_\mathrm{e}=\hbar c(3\pi^2n_\mathrm{e})^{1/3}$, in which $\hbar$ is the reduced Planck constant and $V$ the
volume.

The final pressure constraint is obtained using $E_F=37\MeV$ and by varying the parameters $\ns$, $B$, $K$, and $\esym$ within specific ranges. 
For the parameters describing symmetric nuclear matter, we use the values consistent with the FOPI analysis given by $\ns=0.16\fmiq$, $B=16\MeV$, and a Gaussian distribution for $K$ with $K=200 \pm 25\MeV$ at 1$\sigma$. 
Regarding $\esym$, we apply a uniform prior in the range 31-34$\MeV$. 
We further use results for the pressure of symmetric nuclear matter deduced elsewhere\cite{Dani02Esymm} and disregard all parameter sets, which lead to a pressure that is not consistent with their constraint in the overlapping density range for which ASY-EOS remains sensitive, between 2 and 3$\ns$ (Extended Data Fig.~\ref{fig:AsyEOS_pSNM}).
We note that the value of $K$ has very little influence on the observables measured by ASY-EOS to extract the symmetry energy\cite{Cozma:2017bre}. 
We have explicitly checked the robustness of our results when using larger uncertainty ranges for all nuclear matter parameters in agreement with theoretical predictions\cite{Huth_2021} and found their influence on our final result to be negligible (Extended Data Tab.~\ref{tab:inflated_errors}). 
In particular, we have used a larger range for $S_0$ between 30 and 35 $\MeV$ and the following Gaussian distributions for $\ns$, $B$ and $K$: $\ns=0.164\pm 0.007 \fmiq$, $B=15.86\pm0.57$~MeV, and $K=215\pm40$~MeV at the 1$\sigma$ level.

\noindent
\textbf{Combination of the astronomical and HIC constraints}

To combine the experimental and observational EOS constraints, we use Bayesian inference.
The EOS posterior is given by
\begin{equation}
\begin{aligned}
    p(\textrm{EOS} | \textrm{MMA, HIC}) &\propto p(\textrm{HIC} | \textrm{EOS})\\
    &\times p(\textrm{MMA} | \textrm{EOS}) p(\textrm{EOS})\\
    &= p(\textrm{HIC} | \textrm{EOS}) p(\textrm{EOS} | \textrm{MMA})\\
    &\equiv \mathcal{L}_{\rm{HIC}}(\textrm{EOS})\mathcal{P}_{\rm{MMA}}(\textrm{EOS}),
\end{aligned}
\end{equation}
in which MMA denotes multi-messenger astrophysics, $\mathcal{L}_{\rm{HIC}}(\textrm{EOS})$ is the likelihood of the HIC measurements for a given EOS, and $\mathcal{P}_{\rm{MMA}}(\textrm{EOS})$ is the posterior probability distribution on the EOS based on the multi-messenger observations, which acts as prior for this analysis.
From the HIC experiments, we obtain a posterior of the pressure at a given density, $p(P|n, \rm{HIC})$. 
By combining this with the distribution of probed densities  from the neutrons-over-charged particles sensitivity curve\cite{Russotto:2016ucm}, $p(n |\rm{HIC})$, the joint posterior $p(n, P | {\rm HIC}) = p(P|n, {\rm HIC}) p(n | {\rm HIC})$ is obtained. 
Therefore, the relative faithfulness of the experimental results at various densities is accounted for.
The likelihood $\mathcal{L}_{\rm{HIC}}(\textrm{EOS})$ is given by
\begin{equation}
    \begin{aligned}
    \mathcal{L}_\textrm{HIC}(\textrm{EOS}) &= \int dn\, dP\ p(\textrm{HIC} | n, P)p(n, P |\textrm{EOS})\\
    &\propto \int dn\, dP\ p(n, P | \textrm{HIC})p(n, P |\textrm{EOS})\\
	&\propto \int dn\, dP\ p(n, P | \textrm{HIC})\delta(P-P(n,{\rm EOS}))\\
	&= \int dn\, P(n, P=P(n; {\rm EOS} )| \textrm{HIC})\,,
    \end{aligned}
\end{equation}
in which we used the pressure as a function of density for a given EOS.

\section*{\large Data Availability}
The datasets generated and/or analysed during the current study are available from the corresponding author and on Zenodo (\url{https://doi.org/10.5281/zenodo.6092717}).
The GW data strain that we have analysed in this work was obtained from the Gravitational Wave Open Science Center (ref.\cite{Vallisneri:2014vxa} at \url{https://www.gw-openscience.org}), and the NICER data were obtained from Zenodo (\url{https://doi.org/10.5281/zenodo.3473466},\url{https://doi.org/10.5281/zenodo.4670689} and \url{https://doi.org/10.5281/zenodo.4697625}). 

\section*{\large Code Availability}
All GW models used are implemented in the publicly available software LALSuite (\url{https://git.ligo.org/lscsoft}). 
The bilby and parallel bilby software packages are available at \url{https://git.ligo.org/lscsoft/bilby} and \url{https://git.ligo.org/lscsoft/parallel_bilby}, respectively. 
The gwemlightcurve software is available at \url{https://gwemlightcurves.github.io/}.
The exact code versions of bilby, parallel bilby, LALSuite, and gwemlightcurve that we have used for this work are also available (ref.\cite{dietrich_tim_2020_4114141}).  

\section*{\large Acknowledgments}
We thank Dan Cozma, Christoph Hartnack, Philippe Landry, Paolo Russotto, Yongjia Wang, and LIGO's extreme matter group for valuable comments and useful discussions.

\noindent
\textbf{Funding:} This work was supported by the Deutsche Forschungsgemeinschaft (DFG, German Research Foundation) -- Project-ID 279384907 -- SFB 1245 (SH, AS); the research program of the Netherlands Organization for Scientific Research (NWO) (PTHP, CVDB);
by the U.S. Department of Energy, Office of Science, Office of Nuclear Physics, under contract No.~DE-AC52-06NA25396, by the Laboratory Directed Research and Development program of Los Alamos National Laboratory under project numbers 20190617PRD1 and 20190021DR, and by the U.S. Department of Energy, Office of Science, Office of Advanced Scientific Computing Research, Scientific Discovery through Advanced Computing (SciDAC) program (IT); and by the Max Planck Society (TD).
MB acknowledges support from the Swedish Research Council (Reg. no. 2020-03330) and MWC by the National Science Foundation with grant numbers PHY-2010970 and OAC-2117997.
ALF and WT acknowledge the support of the French-German Collaboration Agreement between IN2P3 - DSM/CEA and GSI. KA acknowledges the support from the Bundesministerium für Bildung und Forschung (BMBF, German Federal Ministry of Education and Research) -- Project-ID 05P19VTFC1 and Helmholtz Graduate School for Hadron and Ion Research (HGS-HIRe).
Computations were performed on the national supercomputer Hawk at the High Performance Computing Center Stuttgart (HLRS) under the grant number 44189 and on SuperMUC-NG at Leibniz Supercomputing Centre Munich under project number pn29ba. 
In addition, computational resources have also been provided by the Los Alamos National Laboratory Institutional Computing Program, which is supported by the U.S. Department of Energy National Nuclear Security Administration under Contract No.~89233218CNA000001, and by the National Energy Research Scientific Computing Center (NERSC), which is supported by the U.S. Department of Energy, Office of Science, under contract No.~DE-AC02-05CH11231.
This research has made use of data, software and/or web tools obtained from the Gravitational
Wave Open Science Center (https://www.gw-openscience.org), a service of LIGO Laboratory, the 
LIGO Scientific Collaboration and the Virgo Collaboration. LIGO is funded by the U.S. National 
Science Foundation. Virgo is funded by the French Centre National de Recherche Scientifique (CNRS), 
the Italian Istituto Nazionale della Fisica Nucleare (INFN) and the Dutch Nikhef, with contributions 
by Polish and Hungarian institutes.

\noindent
\textbf{Authors contributions:} S.~Huth and P.T.H.~Pang share first authorship.\\ 
%https://journals.biologists.com/dev/pages/author-contributions
%Authors ordered according to author list, explanation given below points
%-----
Conceptualization: SH, PTHP, IT, TD, ALF, AS;
%Ideas; formulation or evolution of overarching research goals and aims.
Methodology: SH, PTHP, IT, TD, ALF, AS, MB, MWC;
%Development or design of methodology; creation of models
Data curation: SH, PTHP, IT, TD, MWC;
%Management activities to annotate (produce metadata), scrub data and maintain research data (including software code, where it is necessary for interpreting the data itself) for initial use and later reuse.
Software: SH, PTHP, IT, TD, ALF, MB, MWC;
%Programming, software development; designing computer programs; implementation of the computer code and supporting algorithms; testing of existing code components.
Validation: SH, PTHP, IT, TD, ALF, AS, WT;
%Verification, whether as a part of the activity or separate, of the overall replication/reproducibility of results/experiments and other research outputs.
Formal analysis: SH, PTHP, IT, TD;
%Application of statistical, mathematical, computational, or other formal techniques to analyse or synthesize study data.
Resources: IT, TD, ALF, AS;
%Provision of study materials, reagents, materials, patients, laboratory samples, animals, instrumentation, computing resources, or other analysis tools.
Funding acquisition: IT, TD, ALF, AS, CVDB;
%Acquisition of the financial support for the project leading to this publication.
Project administration: IT, TD, AS;
%Management and coordination responsibility for the research activity planning and execution.
Supervision: IT, TD, ALF, AS, CVDB;
%Oversight and leadership responsibility for the research activity planning and execution, including mentorship external to the core team.
Visualization: SH, PTHP;
%Preparation, creation and/or presentation of the published work, specifically visualization/data presentation.
Writing--original draft: SH, PTHP, IT, TD, ALF, AS, WT, KA;
%Creation and/or presentation of the published work, specifically writing the initial draft (including substantive translation).
Writing--review and editing: SH, PTHP, IT, TD, ALF, AS, WT, KA.
%Preparation, creation and/or presentation of the published work by those from the original research group, specifically critical review, commentary or revision – including pre- or post-publication stages.

\noindent
\textbf{Competing interests:} The authors declare no competing interests.

%Papers containing Supplementary Information contain a statement: Supplementary Information is available for this paper.
\noindent
\textbf{Additional Information:}
Supplementary Information is available for this paper.
Correspondence and requests for materials should be addressed to Sabrina Huth and Peter T.H. Pang.
Reprints and permissions information is available at \url{www.nature.com/reprints}.

\bibliography{refs}
\bibliographystyle{naturemag}

%\newpage
%\clearpage

\renewcommand{\figurename}{Extended Data Figure}
\renewcommand\thefigure{\arabic{figure}}
\setcounter{figure}{0} 
\renewcommand{\tablename}{Extended Data Table}
\renewcommand\thetable{\arabic{table}}
\setcounter{table}{0} 

\begin{figure}[h!]
    \includegraphics[width=\columnwidth]{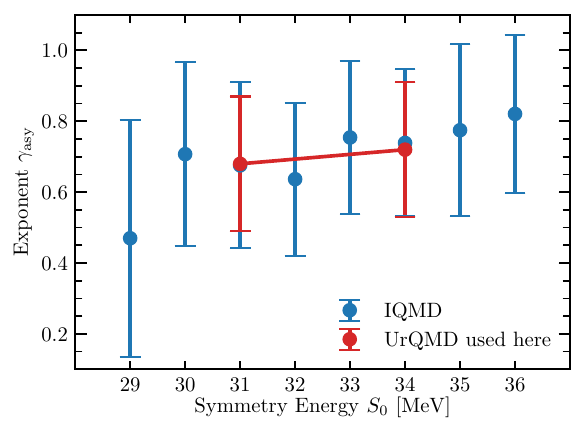}
    \caption{{\bf Constraints on ${\bf \gasy}$ versus symmetry energy ${\bf S_0}$ from two Quantum Molecular Dynamics models.}\label{fig:S0_gammaasy}
    We show the exponent $\gasy$ of the density dependence of the potential part of the symmetry energy, see Eq.~\eqref{eq:Esym_QMD}, as deduced from the analysis of ASY-EOS experimental data using the UrQMD model used in this work\protect\cite{Russotto:2016ucm} (red points) and new simulations from the IQMD model (blue points). 
    The red line indicates the mean value for $\gasy$ along the linear interpolation for the chosen range of $S_0$. Overall, the models are in good agreement with each other and the results suggest that a linear interpolation is reasonable.}
\end{figure}

\begin{figure}[t!]
    \includegraphics[width=\columnwidth]{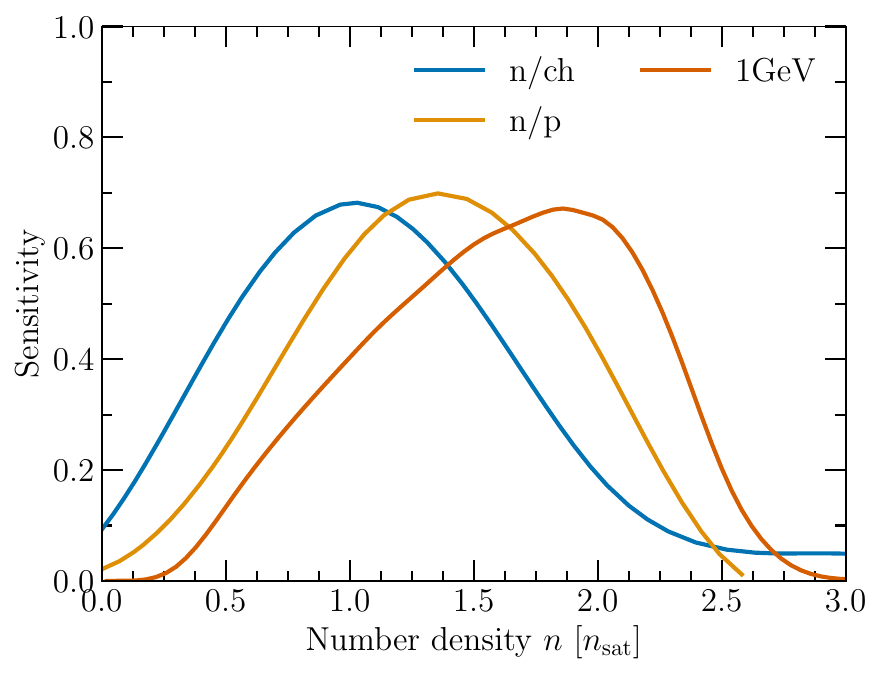}
    \caption{{\bf Comparison between different sensitivity curves.} 
    We show three sensitivity-to-density curves for different observables and incident energies. In particular, the neutron-over-charged-particle (n/ch, used here) and the neutron-over-proton (n/p) sensitivity curves for 400 MeV/nucleon incident energy from Russotto {\it et al.}\protect\cite{Russotto:2016ucm} are compared with the density curve reported by Le F{\`e}vre {\it et al.}\protect\cite{Fevre:2015fza} for the sensitivity of the elliptic flow of protons in Au+Au collisions at 1 GeV/nucleon.}
    \label{fig:sensitivity_curves}
\end{figure}

\begin{figure*}[t!]
    \centering
    \includegraphics[width=\textwidth]{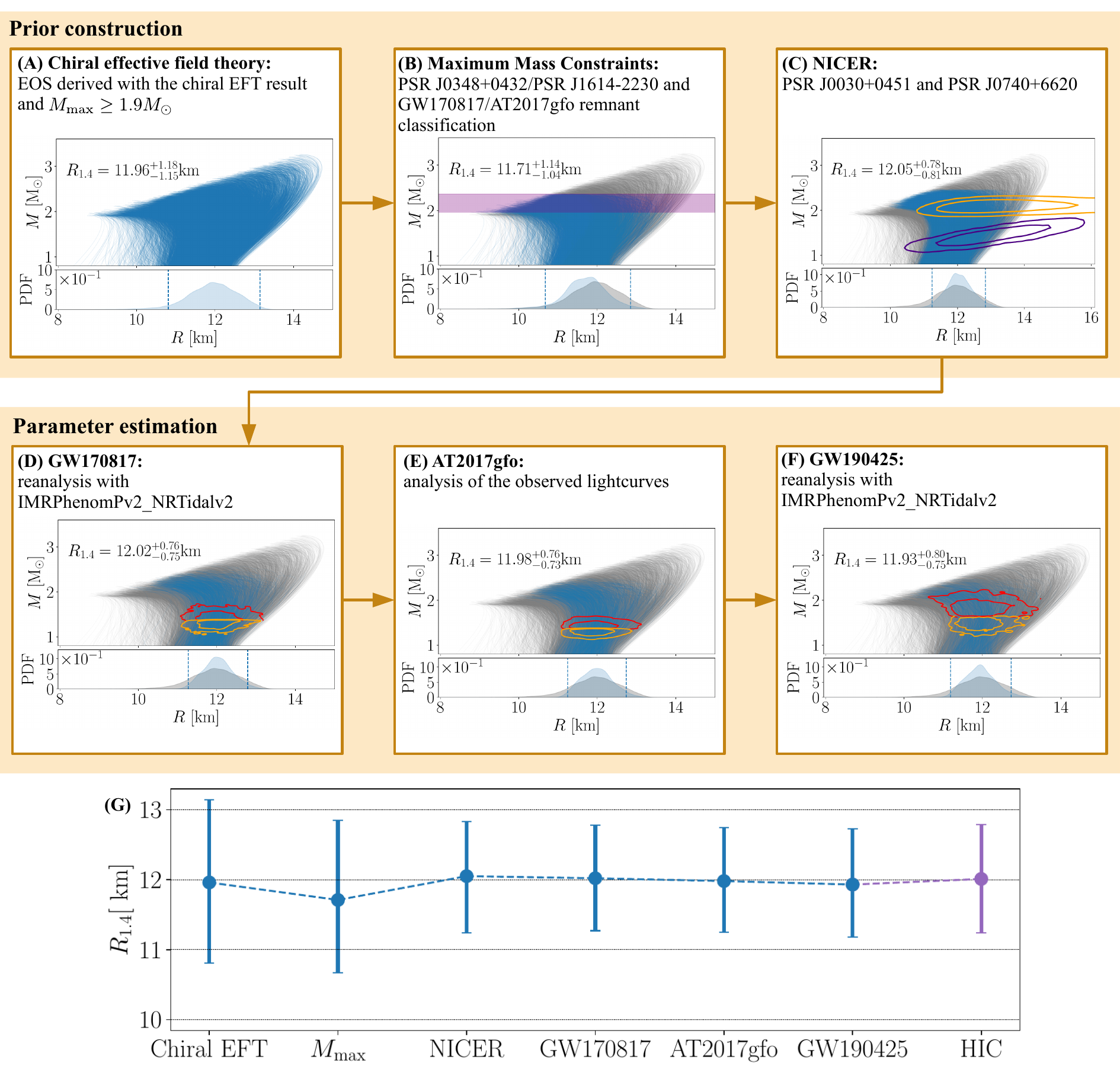}
    \caption{{\bf Constraint on the neutron-star mass and radius with successive astrophysics information.}
    In each panel (except for panel A), EOSs within (outside of) 95\% credible interval are shown as blue (grey) lines. 
    Lower panels indicate the probability distribution function (PDF) for the radius of a 1.4$M_{\odot}$ neutron star, with the 95\% confidence range indicated by dashed lines, in panels (B)-(F) the prior from panel (A) is shown in grey.
    (A) The EOS prior set constrained by chiral EFT calculations up to $1.5n_{\rm sat}$ and $M_{\rm max} \geq 1.9M_\odot$. 
    (B) The EOS set restricted by incorporating information from mass measurements of PSR~J0348+0432, PSR~J1614-2230, and the maximum-mass constraints obtained from GW170817/AT2017gfo. 
    The 95\% confidence interval of the maximum mass posterior probability distribution is shown by the purple band.
    (C) The EOS set further restricted by the NICER mass-radius measurement of PSR~J0030+0451 (purple contours at 68\% and 95\% confidence) and PSR~J0740+6620 (orange contours at 68\% and 95\% confidence). 
    Note that the latter shows the average of the results obtained by Miller \textit{et al.}\protect\cite{Miller:2021qha} and Riley \textit{et al.}\protect\cite{Riley:2021pdl}
    (D) Further restrictions on the EOS set from a reanalysis of the GW170817 using Bayesian inference. 
    Contours at 68\% and 95\% confidence show the mass-radius measurements of the primary (red) and secondary (orange) neutron stars.
    (E) We use the chirp mass, mass ratio, and the EOSs as Bayesian prior for our analysis of AT2017gfo. 
    (F) Further restrictions by analysing GW190425. Again, contours at 68\% and 95\% confidence show the mass-radius measurements of the primary (red) and secondary (orange) neutron stars.
    (G) The radius constraint at each step of this analysis with 95\% confidence ranges. The radius constraint after including HIC experimental data is also shown.}
    \label{fig:astro_steps}
\end{figure*}

\begin{figure}
    \centering
    \includegraphics[width=\columnwidth]{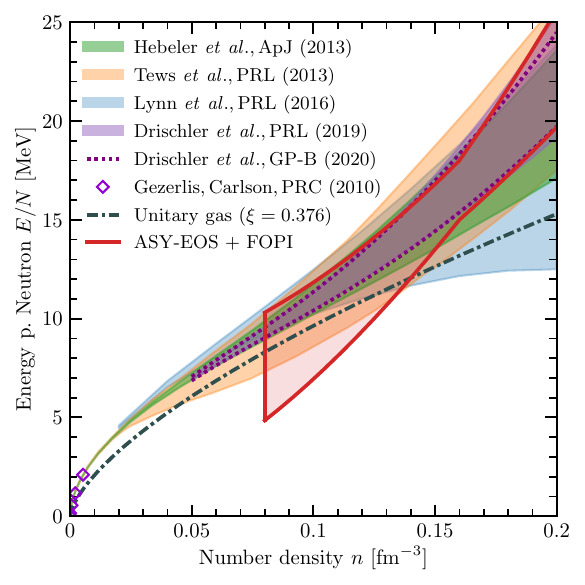}
    \caption{{\bf Constraints for pure neutron matter.}
    Energy per particle $E/N$ of neutron matter as a function of density $n$ for various many-body calculations using chiral EFT interactions from Hebeler {\it et al}.\protect\cite{Hebeler:2013nza}, Tews {\it et al}.\protect\cite{Tews:2012fj}, Lynn {\it et al}.~(used here)\protect\cite{Lynn:2016}, Drischler {\it et al}. PRL\protect\cite{Drischler:2017wtt} and GP-B\protect\cite{Dris20GPB}, and low-density quantum Monte Carlo results from Gezerlis and Carlson\protect\cite{Geze09neutmat}. Overall, the results from these calculations are in good agreement with each other.
    We also show the energy per particle of a unitary Fermi gas of neutrons, which has been proposed as a lower bound for the energy of neutron matter\protect\cite{Tews17NMatter}. 
    Finally, we compare the theoretical results with the constraint from the ASY-EOS and FOPI experiments (red), which is used as a constraint for neutron matter in the main work.}
    \label{fig:PNM_constraints}
\end{figure}

\clearpage

\begin{figure}
    \centering
    \includegraphics[width=\columnwidth]{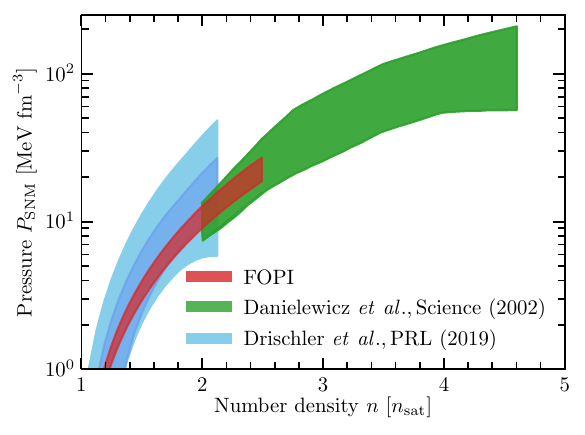}
    \caption{{\bf Comparison of the pressure of symmetric nuclear matter for experiment and theory.} 
    The pressure band from the FOPI experiment\protect\cite{Fevre:2015fza} at the $1\sigma$ level (red) for the incompressibility is consistent with the chiral EFT constraint from Drischler {\it et al.}\protect\cite{Drischler:2017wtt,Leon19fRGeos} at N$^2$LO (light blue) and N$^3$LO (dark blue). 
    The experimental uncertainty band is smaller than the theoretical one because the empirical saturation point used for extracting the experimental results has smaller uncertainties compared to theoretical estimates from chiral EFT. 
    Between 2-3$\ns$, we additionally constrain the FOPI results with the constraint from Danielewicz {\it et al.}\protect\cite{Dani02Esymm} (green), which has no statistical interpretation. This excludes the highest values for the incompressibility $K$ from the FOPI distribution and also influences symmetric matter at smaller densities, which depends on the range of $K$.
    However, both constraints are in very good agreement with each other and the impact of the additional Danielewicz {\it et al.} constraint is small in our analysis. 
    }
    \label{fig:AsyEOS_pSNM}
\end{figure}

\begin{table*}
    \begin{center}
    \renewcommand{\arraystretch}{1.3}
    \caption{{Impact of the EOS prior: Maximum density of chiral EFT and of prior distribution of $R_{1.4}$.}}\label{tab:prior_table}
    \begin{tabular}{|m{0.09\columnwidth}||>{\centering}m{0.165\columnwidth}>{\centering}m{0.165\columnwidth}>{\centering}m{0.165\columnwidth}|>{\centering}m{0.165\columnwidth}>{\centering}m{0.165\columnwidth}>{\centering\arraybackslash}m{0.165\columnwidth}|}
    \hline
    & \multicolumn{6}{c|}{Natural prior on $R_{1.4}$}\\
    \hline
    & \multicolumn{3}{c|}{Chiral EFT up to $1.5 n_{\rm sat}$} & \multicolumn{3}{c|}{Chiral EFT up to $1 n_{\rm sat}$} \\
    $P/R$ & HIC only & Astro only & Astro+HIC & HIC only & Astro only& Astro+HIC \\
    \hline\hline
    $1.0 n_{\rm sat}$ & $2.05^{+0.49}_{-0.45}$ & $2.00^{+0.52}_{-0.49}$ & $2.11^{+0.49}_{-0.52}$ & $1.95^{+0.51}_{-0.39}$ & $1.87^{+0.51}_{-0.41}$ & $1.95^{+0.50}_{-0.43}$ \\
    $1.5 n_{\rm sat}$ & $6.06^{+1.85}_{-2.04}$ & $5.84^{+1.96}_{-2.26}$ & $6.25^{+1.90}_{-2.26}$ & $10.77^{+29.80}_{-8.81}$ & $8.98^{+8.41}_{-4.30}$ & $9.12^{+6.66}_{-4.36}$ \\
    $2.0 n_{\rm sat}$ & $19.47^{+33.63}_{-11.67}$ & $18.44^{+16.24}_{-9.69}$ & $19.07^{+15.27}_{-10.53}$ & $33.02^{+76.25}_{-31.06}$ & $26.11^{+24.36}_{-17.81}$ & $26.21^{+21.85}_{-17.16}$ \\
    $2.5 n_{\rm sat}$ & $47.78^{+75.96}_{-32.96}$ & $45.05^{+39.80}_{-19.62}$ & $45.43^{+40.41}_{-19.11}$ & $68.31^{+114.74}_{-66.35}$ & $54.19^{+38.50}_{-20.67}$ & $54.33^{+35.54}_{-21.69}$ \\
    \hline\hline
    $1.0M_{\odot}$ & $11.89^{+0.79}_{-0.98}$ & $11.76^{+0.65}_{-0.71}$ & $11.88^{+0.57}_{-0.76}$ & $12.68^{+1.44}_{-1.41}$ & $12.36^{+0.95}_{-0.90}$ & $12.40^{+0.85}_{-0.89}$  \\
    $1.4M_{\odot}$ & $12.06^{+1.13}_{-1.18}$ & $11.94^{+0.79}_{-0.78}$ & $12.01^{+0.78}_{-0.77}$ & $12.96^{+1.87}_{-1.84}$ & $12.53^{+1.22}_{-1.03}$ & $12.56^{+1.07}_{-1.01}$  \\
    $1.6M_{\odot}$ & $12.11^{+1.33}_{-1.33}$ & $11.98^{+0.93}_{-0.79}$ & $12.03^{+0.98}_{-0.75}$ & $13.05^{+2.11}_{-2.08}$ & $12.55^{+1.31}_{-1.10}$ & $12.57^{+1.22}_{-1.04}$  \\
    $2.0M_{\odot}$ & $12.19^{+1.71}_{-1.59}$ & $11.88^{+1.23}_{-1.10}$ & $11.91^{+1.24}_{-1.11}$ & $13.21^{+2.53}_{-2.38}$ & $12.32^{+1.58}_{-1.49}$ & $12.33^{+1.56}_{-1.44}$ \\
    \hline
    \end{tabular}
    \\
    \begin{tabular}{|m{0.09\columnwidth}||>{\centering}m{0.165\columnwidth}>{\centering}m{0.165\columnwidth}>{\centering}m{0.165\columnwidth}|>{\centering}m{0.165\columnwidth}>{\centering}m{0.165\columnwidth}>{\centering\arraybackslash}m{0.165\columnwidth}|}
    \hline
    & \multicolumn{6}{c|}{Uniform prior on $R_{1.4}$}\\
    \hline
    & \multicolumn{3}{c|}{Chiral EFT up to $1.5 n_{\rm sat}$} & \multicolumn{3}{c|}{Chiral EFT up to $1 n_{\rm sat}$} \\
    $P/R$ & HIC only & Astro only & Astro+HIC & HIC only & Astro only& Astro+HIC \\
    \hline\hline
    $1.0 n_{\rm sat}$ & $2.05^{+0.46}_{-0.54}$ & $1.92^{+0.64}_{-0.45}$ & $2.18^{+0.43}_{-0.68}$ & $1.98^{+0.49}_{-0.40}$ & $1.90^{+0.52}_{-0.43}$ & $2.00^{+0.49}_{-0.46}$ \\
    $1.5 n_{\rm sat}$ & $6.12^{+1.75}_{-2.43}$ & $5.56^{+2.45}_{-2.15}$ & $6.57^{+1.66}_{-2.92}$ & $9.11^{+42.6}_{-7.53}$ & $8.22^{+6.51}_{-5.53}$ & $8.58^{+6.62}_{-5.70}$ \\
    $2.0 n_{\rm sat}$ & $17.04^{+46.81}_{-12.56}$ & $18.19^{+27.15}_{-12.37}$ & $19.93^{+29.61}_{-12.96}$ & $23.84^{+100.12}_{-22.25}$ & $22.56^{+21.12}_{-18.76}$ & $23.45^{+21.97}_{-18.10}$\\
    $2.5 n_{\rm sat}$ & $38.39^{+98.48}_{-34.37}$ & $44.28^{+47.06}_{-24.88}$ & $47.03^{+52.26}_{-22.44}$ & $48.34^{+154.87}_{-46.75}$ & $46.39^{+38.20}_{-31.12}$ & $47.89^{+37.10}_{-32.47}$\\
    \hline\hline
    $1.0M_{\odot}$ & $11.70^{+1.25}_{-2.23}$ & $11.72^{+0.91}_{-0.89}$ & $11.96^{+0.78}_{-1.02}$ & $12.27^{+1.92}_{-3.01}$ & $12.15^{+1.07}_{-1.39}$ & $12.25^{+1.04}_{-1.41}$ \\
    $1.4M_{\odot}$ & $11.81^{+1.62}_{-2.30}$ & $11.90^{+1.18}_{-0.92}$ & $12.08^{+1.18}_{-0.94}$ & $12.32^{+2.60}_{-2.89}$ & $12.22^{+1.31}_{-1.42}$ & $12.33^{+1.26}_{-1.52}$ \\
    $1.6M_{\odot}$ & $11.81^{+1.86}_{-2.33}$ & $11.94^{+1.37}_{-0.96}$ & $12.10^{+1.34}_{-1.02}$ & $12.29^{+2.93}_{-2.87}$ & $12.20^{+1.44}_{-1.43}$ & $12.30^{+1.42}_{-1.50}$ \\
    $2.0M_{\odot}$ & $12.37^{+1.82}_{-2.69}$ & $11.82^{+1.71}_{-1.27}$ & $11.97^{+1.80}_{-1.27}$ & $12.92^{+3.04}_{-3.22}$ & $11.88^{+1.85}_{-1.57}$ & $11.94^{+1.85}_{-1.59}$\\
    \hline
    \end{tabular}
    \end{center}
    Comparison of the 95\% credible interval for the pressure $[\MeV \fmiq]$ and radius [km] of neutron stars when including only HIC experiments, only astrophysical observations, or the combined HIC and astrophysics results for chiral EFT constraints up to $1.5\ns$ and up to $1\ns$, and for using a natural and uniform prior on $R_{1.4}$.
    We find that differences for pressures and neutron-star radii are small between both prior choices when Astro+HIC data constraints are employed.
    Applying constraints from chiral EFT only up to $1\ns$ allows for a broader and stiffer EOS prior at higher densities since information up to $1.5\ns$ is discarded. As a consequence, the EOSs including HIC only and to a lesser extent the combination of HIC and observational constraints become stiffer leading to an increase of neutron-star radii. This effect is larger when using a natural instead of a uniform prior in radius. Nevertheless, the impact for the natural prior is only around 5\%.
\end{table*}

\begin{table*}
    \begin{center} 
    \renewcommand{\arraystretch}{1.3}
    \caption{{Impact of EOS extension scheme.}}\label{tab:cse_pwp_comparison}
    \begin{tabular}{|m{0.09\columnwidth}||>{\centering} m{0.165\columnwidth}>{\centering} m{0.165\columnwidth}>{\centering}m{0.165\columnwidth}|>{\centering}m{0.165\columnwidth}>{\centering\arraybackslash}m{0.165\columnwidth}>{\centering\arraybackslash}m{0.165\columnwidth}|}
    \hline
    & \multicolumn{3}{c|}{Speed-of-sound extension} & \multicolumn{3}{c|}{Piecewise-polytrope extension} \\
    $P/R$ & HIC only & Astro only & Astro+HIC & HIC only & Astro only & Astro+HIC\\
    \hline\hline
    $1.0 n_{\rm sat}$ & $2.05^{+0.49}_{-0.45}$ & $2.00^{+0.52}_{-0.49}$ & $2.11^{+0.49}_{-0.52}$ & $2.06^{+0.49}_{-0.44}$ & $1.96^{+0.54}_{-0.45}$ & $2.10^{+0.49}_{-0.51}$ \\
    $1.5 n_{\rm sat}$ & $6.06^{+1.85}_{-2.04}$ & $5.84^{1.96}_{-2.26}$ & $6.25^{+1.90}_{-2.26}$ & $6.06^{+1.85}_{-1.96}$ & $5.66^{+2.15}_{-2.00}$ & $6.20^{+1.93}_{-2.17}$ \\
    $2.0 n_{\rm sat}$ & $19.47^{+33.63}_{-11.67}$ & $18.44^{+16.24}_{-9.69}$ & $19.07^{+15.27}_{-10.53}$ & $19.00^{+17.6}_{-8.34}$ & $18.96^{+15.40}_{-8.40}$ & $19.64^{+15.83}_{-8.60}$ \\
    $2.5 n_{\rm sat}$ & $47.78^{+75.96}_{-32.96}$ & $45.05^{+39.80}_{-19.62}$ & $45.43^{+40.41}_{-19.11}$ & $43.72^{+39.81}_{-18.98}$ & $44.77^{+35.36}_{-18.86}$ & $45.27^{+36.77}_{-18.00}$ \\
    \hline\hline
    $1.0M_{\odot}$ & $11.89^{+0.79}_{-0.98}$ & $11.76^{+0.65}_{-0.71}$ & $11.88^{+0.57}_{-0.76}$ & $11.90^{+0.74}_{-0.89}$ & $11.80^{+0.70}_{-0.69}$ & $11.92^{+0.67}_{-0.71}$ \\
    $1.4M_{\odot}$ & $12.06^{+1.13}_{-1.18}$ & $11.94^{+0.79}_{-0.78}$ & $12.01^{+0.78}_{-0.77}$ & $12.02^{+0.96}_{-1.01}$ & $11.97^{+0.84}_{-0.77}$ & $12.05^{+0.83}_{-0.79}$ \\
    $1.6M_{\odot}$ & $12.11^{+1.33}_{-1.33}$ & $11.98^{+0.93}_{-0.79}$ & $12.03^{+0.98}_{-0.75}$ & $12.05^{+1.11}_{-1.11}$ & $12.01^{+0.94}_{-0.83}$ & $12.07^{+0.95}_{-0.84}$ \\
    $2.0M_{\odot}$ & $12.19^{+1.71}_{-1.59}$ & $11.88^{+1.23}_{-1.10}$ & $11.91^{+1.24}_{-1.11}$ & $12.02^{+1.35}_{-1.39}$ & $11.88^{+1.22}_{-1.11}$ & $11.92^{+1.32}_{-1.08}$ \\
    \hline
    \end{tabular}
    \end{center}
    Comparison of the 95\% credible interval for the pressure $[\MeV \fmiq]$ and radius [km] of neutron stars when including only HIC results, only astrophysical observations, and for combined HIC and astrophysics results for different EOS extension schemes used. The piecewise-polytrope scheme extends the EOS beyond $1.5\ns$ with five polytropic segments with randomly chosen transition densities and polytropic indices. The differences of the pressure estimates between the two EOS extension schemes is less than $3\%$ and the difference between the radius estimates is less than $0.5\%$ for the combined HIC and astrophysics results.
\end{table*}

\begin{table}[t!]
    \begin{center}
    \renewcommand{\arraystretch}{1.3}
    \caption{{Sensitivity on uncertainties of nuclear matter properties.}}\label{tab:inflated_errors}
    \begin{tabular}{|m{0.09\columnwidth}||>{\centering} m{0.165\columnwidth}>{\centering}m{0.165\columnwidth}|>{\centering}m{0.165\columnwidth}>{\centering\arraybackslash}m{0.165\columnwidth}|}
    \hline
    & \multicolumn{2}{c|}{HIC parameters} & \multicolumn{2}{c|}{Enlarged variations} \\
    $P/R$ & HIC only & Astro+HIC & HIC only & Astro+HIC\\
    \hline\hline
    $1.0 n_{\rm sat}$ & $2.05^{+0.49}_{-0.45} $ & $ 2.11^{+0.49}_{-0.52}$ & $2.05^{+0.50}_{-0.45} $ & $ 2.09^{+0.47}_{-0.52}$ \\
    $1.5 n_{\rm sat}$ & $6.06^{+1.85}_{-2.04} $ & $ 6.25^{+1.90}_{-2.26}$ & $6.00^{+1.90}_{-2.00} $ & $ 6.18^{+1.88}_{-2.25}$ \\
    $2.0 n_{\rm sat}$ & $19.47^{+33.63}_{-11.67} $ & $19.07^{+15.27}_{-10.53}$ & $19.34^{+35.65}_{-11.54} $ & $18.98^{+14.97}_{-9.92}$ \\
    $2.5 n_{\rm sat}$ & $47.78^{+75.96}_{-32.96} $ & $45.43^{+40.41}_{-19.11}$ & $47.36^{+81.44}_{-28.09} $ & $45.49^{+40.05}_{-20.58}$ \\
    \hline\hline
    $1.0M_{\odot}$ & $11.89^{+0.79}_{-0.98}$ & $ 11.88^{+0.57}_{-0.76}$ & $11.87^{+0.81}_{-0.97} $ & $11.86^{+0.58}_{-0.78}$ \\
    $1.4M_{\odot}$ & $12.06^{+1.13}_{-1.18}$ & $ 12.01^{+0.78}_{-0.77}$ & $12.05^{+1.12}_{-1.20}$ & $ 12.00^{+0.75}_{-0.80}$ \\
    $1.6M_{\odot}$ & $12.11^{+1.33}_{-1.33}$ & $ 12.03^{+0.98}_{-0.75}$ & $12.10^{+1.35}_{-1.32}$ & $ 12.03^{+0.92}_{-0.80}$ \\
    $2.0M_{\odot}$ & $12.19^{+1.71}_{-1.59}$ & $ 11.91^{+1.24}_{-1.11}$ & $12.17^{+1.70}_{-1.62}$ & $ 11.91^{+1.23}_{-1.15}$ \\
    \hline
    \end{tabular}
    \end{center}
    Comparison of the 95\% credible interval for the pressure $[\MeV \fmiq]$ and radius [km] of neutron stars when using ranges for nuclear matter properties as published for the FOPI and ASY-EOS experiments\protect\cite{Russotto:2016ucm, Fevre:2015fza} and when inflating the uncertainties according to theoretical calculations. 
    We present results when including only information from HIC experiments and for the combined HIC and astrophysics information.
    In particular, we extend the range for the symmetry energy at saturation density to $S_0=30-35\MeV$ by extrapolating $\gasy$ linearly. We use Gaussian distributions for $\ns$, $B$, and $K$ describing symmetric nuclear matter and vary these parameters within their empirical ranges (at $1\sigma$): $\ns = 0.164 \pm 0.007\fmiq$, $B=15.86 \pm 0.57\MeV$\protect\cite{Drischler:2017wtt} and $K=215 \pm 40\MeV$ from microscopic calculations\protect\cite{Drischler:2017wtt,Hebe11fits, Dris16asym}, which is in good agreement with the FOPI results.
    \end{table}

\begin{table*}
    \begin{center} 
    \renewcommand{\arraystretch}{1.3}
    \caption{{Impact of the sensitivity curve for the ASY-EOS experiment.}}\label{tab:sensitivity_table}
    \begin{tabular}{|m{0.09\columnwidth}||>{\centering} m{0.165\columnwidth}>{\centering}m{0.165\columnwidth}|>{\centering} m{0.165\columnwidth}>{\centering}m{0.165\columnwidth}|>{\centering}m{0.165\columnwidth}>{\centering\arraybackslash}m{0.165\columnwidth}|}
    \hline
    & \multicolumn{2}{c|}{n/ch sensitivity} & \multicolumn{2}{c|}{n/p sensitivity} & \multicolumn{2}{c|}{Window 1-2$\ns$} \\
    $P/R$ & HIC only & Astro+HIC & HIC only & Astro+HIC & HIC only & Astro+HIC\\
    \hline\hline
    $1.0 n_{\rm sat}$ & $2.05^{+0.49}_{-0.45}$ & $2.11^{+0.49}_{-0.52}$ & $2.10^{+0.45}_{-0.49}$ & $2.13^{+0.46}_{-0.54}$ & $2.23^{+0.32}_{-0.50}$ & $2.28^{+0.35}_{-0.55}$\\
    $1.5 n_{\rm sat}$ & $6.06^{+1.85}_{-2.04}$ & $6.25^{+1.90}_{-2.26}$ & $6.23^{+1.68}_{-2.16}$ & $6.34^{+1.83}_{-2.30}$ & $6.76^{+1.15}_{-2.13}$ & $6.93^{+1.39}_{-2.17}$\\
    $2.0 n_{\rm sat}$ & $19.47^{+33.63}_{-11.67}$ & $19.07^{+15.27}_{-10.53}$ & $19.62^{+33.36}_{-10.81}$ & $19.20^{+15.42}_{-9.21}$ & $21.41^{+30.60}_{-9.02}$ & $20.59^{+16.10}_{-8.36}$\\
    $2.5 n_{\rm sat}$ & $47.78^{+75.96}_{-32.96}$ & $45.43^{+40.41}_{-19.11}$ & $47.61^{+79.33}_{-32.61}$ & $45.62^{+40.81}_{-18.61}$ & $54.71^{+66.27}_{-36.26}$ & $48.60^{+39.47}_{-19.32}$\\
    \hline\hline
    $1.0M_{\odot}$ & $11.89^{+0.79}_{-0.98}$ & $11.88^{+0.57}_{-0.76}$ & $11.92^{+0.78}_{-0.95}$ & $11.91^{+0.61}_{-0.73}$ & $12.09^{+0.59}_{-0.63}$ & $12.06^{+0.48}_{-0.56}$\\
    $1.4M_{\odot}$ & $12.06^{+1.13}_{-1.18}$ & $12.01^{+0.78}_{-0.77}$ & $12.09^{+1.12}_{-1.14}$ & $12.02^{+0.78}_{-0.76}$ & $12.26^{+0.96}_{-0.84}$ & $12.17^{+0.73}_{-0.60}$\\
    $1.6M_{\odot}$ & $12.11^{+1.33}_{-1.33}$ & $12.03^{+0.98}_{-0.75}$ & $12.13^{+1.31}_{-1.30}$ & $12.05^{+0.91}_{-0.79}$ & $12.33^{+1.14}_{-1.05}$ & $12.19^{+0.81}_{-0.76}$\\
    $2.0M_{\odot}$ & $12.19^{+1.71}_{-1.59}$ & $11.91^{+1.24}_{-1.11}$ & $12.20^{+1.68}_{-1.60}$ & $11.91^{+1.25}_{-1.11}$ & $12.42^{+1.44}_{-1.48}$ & $12.06^{+1.14}_{-1.20}$\\
    \hline
    \end{tabular}
    \end{center}
    Comparison of the 95\% credible interval for the pressure $[\MeV \fmiq]$ and radius [km] of neutron stars when including only HIC experiments and for combined HIC and astrophysics results for different sensitivity curves. 
    In particular, we compare our standard results using the neutron over charged particles (n/ch) sensitivity curve\protect\cite{Russotto:2016ucm} with the neutron over proton (n/p) sensitivity from Russotto {\it et al.}\protect\cite{Russotto:2016ucm}, which peaks at $1.5\ns$. 
    We find that our results are robust and differences for both sensitivity curves are small.  
    Additionally, we compare the results to calculations where the ASY-EOS data is implemented using a uniform prior in density between 1-2$\ns$ (labelled Window). 
    For the latter choice, we generally find larger pressures and larger neutron-star radii because the n/ch and n/p sensitivy curves decrease rapidly after their maxima at 1 and $1.5\ns$, lowering the impact of the ASY-EOS constraint at higher densities.
    However, differences for radii and pressures remain small once Astro+HIC data is included.
\end{table*}

\begin{table*}
    \begin{center}
    \caption{Future prospects for the ASY-EOS experiment.}\label{tab:expolratory_table}
    \hspace{-0.7cm}
    \renewcommand{\arraystretch}{1.3}
    \begin{tabular}{|m{0.09\columnwidth}||>{\centering} m{0.165\columnwidth}>{\centering}m{0.165\columnwidth}>{\centering}m{0.165\columnwidth}|>{\centering} m{0.165\columnwidth}>{\centering}m{0.165\columnwidth}|>{\centering}m{0.165\columnwidth}>{\centering\arraybackslash}m{0.165\columnwidth}|>{\centering}m{0.165\columnwidth}>{\centering\arraybackslash}m{0.165\columnwidth}|}
    \hline
    & \multicolumn{3}{c|}{Current setup} & \multicolumn{2}{c|}{1 GeV sensitivity} & \multicolumn{2}{c|}{1 GeV sensitivity and} & \multicolumn{2}{c|}{1 GeV sensitivity with}\\ 
    & \multicolumn{3}{c|}{} & \multicolumn{2}{c|}{} & \multicolumn{2}{c|}{halved uncertainty on HIC} & \multicolumn{2}{c|}{a $1n_{\rm sat}$ lower cutoff}\\
    $P/R$ & HIC only & Astro & Astro+HIC & HIC only & Astro+HIC & HIC only & Astro+HIC & HIC only & Astro+HIC\\
    \hline\hline
    $1.0 n_{\rm sat}$ & $2.05^{+0.49}_{-0.45}$ & $2.00^{+0.52}_{-0.49}$ & $2.11^{+0.49}_{-0.52}$ & $2.10^{+0.49}_{-0.45}$ & $2.13^{+0.47}_{-0.53}$ & $2.12^{+0.43}_{-0.48}$ & $2.16^{+0.43}_{-0.55}$ & $2.07^{+0.48}_{-0.45}$ & $2.12^{+0.47}_{-0.53}$\\
    $1.5 n_{\rm sat}$ & $6.06^{+1.85}_{-2.04}$ & $5.84^{+1.96}_{-2.26}$ & $6.25^{+1.90}_{-2.26}$ & $6.20^{+1.71}_{-2.13}$ & $6.35^{+1.80}_{-2.31}$ & $5.84^{+1.96}_{-2.26}$ & $6.44^{+1.77}_{-2.21}$ & $6.11^{+1.80}_{-2.02}$ & $6.33^{+1.82}_{-2.33}$\\
    $2.0 n_{\rm sat}$ & $19.47^{+33.63}_{-11.67}$ & $18.44^{+16.24}_{-9.69}$ & $19.07^{+15.27}_{-10.53}$ & $19.42^{+28.90}_{-11.69}$ & $19.14^{+14.24}_{-8.97}$ & $19.73^{+29.32}_{-11.49}$ & $19.32^{+13.93}_{-8.74}$ & $18.66^{+22.18}_{-8.65}$ & $18.52^{+11.08}_{-7.04}$ \\
    $2.5 n_{\rm sat}$ & $47.78^{+75.96}_{-32.96}$ & $45.05^{+39.80}_{-19.62}$ & $45.43^{+40.41}_{-19.11}$ & $47.13^{+75.65}_{-27.86}$ & $45.3^{+40.52}_{-17.24}$ & $48.20^{+78.30}_{-24.83}$ & $45.73^{+38.03}_{-18.47}$ & $44.31^{+60.85}_{-21.23}$ & $43.30^{+31.65}_{-15.92}$ \\
    \hline\hline
    $1.0M_{\odot}$ & $11.89^{+0.79}_{-0.98}$ & $11.76^{+0.65}_{-0.71}$ & $11.88^{+0.57}_{-0.76}$ & $11.91^{+0.74}_{-0.93}$ & $11.91^{+0.55}_{-0.76}$ & $11.96^{+0.72}_{-0.85}$ & $11.94^{+0.54}_{-0.71}$ & $11.85^{+0.66}_{-0.78}$ & $11.85^{+0.55}_{-0.70}$  \\
    $1.4M_{\odot}$ & $12.06^{+1.13}_{-1.18}$ & $11.94^{+0.79}_{-0.78}$ & $12.01^{+0.78}_{-0.77}$ & $12.08^{+1.09}_{-1.10}$ & $12.02^{+0.76}_{-0.73}$ & $12.11^{+1.07}_{-1.01}$ & $12.04^{+0.72}_{-0.71}$ & $11.97^{+0.95}_{-0.85}$ & $11.96^{+0.66}_{-0.67}$  \\
    $1.6M_{\odot}$ & $12.11^{+1.33}_{-1.33}$ & $11.98^{+0.93}_{-0.79}$ & $12.03^{+0.98}_{-0.75}$ & $12.12^{+1.30}_{-1.24}$ & $12.04^{+0.88}_{-0.75}$ & $12.15^{+1.26}_{-1.16}$ & $12.06^{+0.85}_{-0.74}$ & $11.99^{+1.14}_{-0.95}$ & $11.96^{+0.79}_{-0.66}$  \\
    $2.0M_{\odot}$ & $12.19^{+1.71}_{-1.59}$ & $11.88^{+1.23}_{-1.10}$ & $11.91^{+1.24}_{-1.11}$ & $12.17^{+1.71}_{-1.51}$ & $11.90^{+1.21}_{-1.10}$ & $12.18^{+1.66}_{-1.46}$ & $11.92^{+1.19}_{-1.07}$ & $11.93^{+1.61}_{-1.32}$ & $11.79^{+1.08}_{-0.95}$  \\
    \hline
    \end{tabular}
    \end{center}
    Comparison of the 95\% credible interval for the pressure $[\MeV \fmiq]$ and the radius [km] of a neutron star when including only HIC experiments and for combined HIC and astrophysics results for different future improvements. In particular, we show the results with the 1 GeV sensitivity curve (see Extended Data  Fig.~\ref{fig:sensitivity_curves}) applied to the current measurement (\textit{second column}), when additionally halving the uncertainty on $\gamma_{\rm asy}$ (\textit{third column}), and when using a lower cutoff density of $1n_{\rm sat}$ instead of $0.5n_{\rm sat}$ (\textit{fourth column}). For all the exploratory setups, HIC data is showing a stronger impact on the EOS constraint than the current setup. The result with a density cutoff is showing a significant decrease in uncertainty compared to the result of this work. Therefore, to achieve a stronger constraint on the EOS, improvements to the low-density part of the HIC constraint will be most important.
\end{table*}

\

\renewcommand{\tablename}{Supplementary Table}
\renewcommand\thetable{\arabic{table}}
\setcounter{table}{0} 

\begin{table}[]
    \begin{center}
    \renewcommand{\arraystretch}{1.3}
    \caption{{Kullback–Leibler divergence with different observation input.}}\label{tab:KL_divergence}
    \begin{tabular}{|m{0.09\columnwidth}||>{\centering} m{0.165\columnwidth}>{\centering}m{0.165\columnwidth}>{\centering\arraybackslash}m{0.165\columnwidth}|}
    \hline
    $D_{\rm KL}$ &  Astro only & HIC only & Astro+HIC\\
    \hline \hline
    $1.0\ns$ & $0.079$ & $0.109$ & $0.270$ \\
    $1.5\ns$ & $0.075$ & $0.108$ & $0.266$ \\
    $2.0\ns$ & $0.112$ & $0.019$ & $0.174$ \\
    $2.5\ns$ & $0.244$ & $0.006$ & $0.274$ \\
    \hline\hline
    $1.0M_{\odot}$ & $0.090$ & $0.054$ & $0.128$ \\
    $1.4M_{\odot}$ & $0.185$ & $0.022$ & $0.210$ \\
    $1.6M_{\odot}$ & $0.225$ & $0.015$ & $0.251$ \\
    $2.0M_{\odot}$ & $0.228$ & $0.008$ & $0.222$ \\
    \hline
    \end{tabular}
    \end{center}
     Comparison of the Kullback–Leibler divergence (KL divergence) $D_{\rm KL}{\rm (posterior | prior)}$ of pressure and radius of a neutron star with respect to the prior in bits when including only astrophysical constraints, only HIC experimental data, and for the combination of both. The KL divergence quantifies the additional information encoded in the posterior distribution with respect to the prior distribution. A KL divergence of zero indicates that the two distributions are identical. For reference, the KL divergence $D_{\rm KL}(\mathcal{N}(0, 1/4)|\mathcal{N}(0, 1)) \approx 1.2{\rm bits}$, where $N(\mu, \sigma^2)$ is a normal distribution with mean $\mu$ and variance $\sigma^2$. The KL divergence for the pressure at $1.0\ns$ and $1.5\ns$, using only HIC experimental input, is higher than that of the result using only astrophysical observations. Therefore, the HIC experiment has higher impact than astrophysical observations for pressures below $1.5\ns$.
\end{table}

\begin{table}
    \begin{center}
    \renewcommand{\arraystretch}{1.3}
    \caption{{Impact of the parameterisation for symmetric nuclear matter.}}\label{tab:skyrme_taylor_table}
    \begin{tabular}{|m{0.09\columnwidth}||>{\centering} m{0.165\columnwidth}>{\centering}m{0.165\columnwidth}|>{\centering}m{0.165\columnwidth}>{\centering\arraybackslash}m{0.165\columnwidth}|}
    \hline
    & \multicolumn{2}{c|}{SNM form used here} & \multicolumn{2}{c|}{Taylor expansion} \\
    $P/R$ & HIC only & Astro+HIC & HIC only & Astro+HIC \\
    \hline\hline
    $1.0 n_{\rm sat}$ & $2.05^{+0.49}_{-0.45}$ & $2.11^{+0.49}_{-0.52}$ & $1.95^{+0.52}_{-0.44}$ & $2.01^{+0.51}_{-0.47}$\\
    $1.5 n_{\rm sat}$ & $6.06^{+1.85}_{-2.04}$ & $6.25^{+1.90}_{-2.26}$ & $5.61^{+2.04}_{-2.00}$ & $5.87^{+1.99}_{-2.14}$\\
    $2.0 n_{\rm sat}$ & $19.47^{+33.63}_{-11.67}$ & $19.07^{+15.27}_{-10.53}$ & $18.80^{+32.63}_{-12.89}$ & $18.72^{+16.57}_{-9.34}$\\
    $2.5 n_{\rm sat}$ & $47.78^{+75.96}_{-32.96}$ & $45.43^{+40.41}_{-19.11}$ & $47.58^{+77.40}_{-31.93}$ & $45.66^{+41.66}_{-19.19}$\\
    \hline\hline
    $1.0M_{\odot}$ & $11.89^{+0.79}_{-0.98}$ & $11.88^{+0.57}_{-0.76}$ & $11.77^{+0.84}_{-0.97}$ & $11.79^{+0.60}_{-0.71}$\\
    $1.4M_{\odot}$ & $12.06^{+1.13}_{-1.18}$ & $12.01^{+0.78}_{-0.77}$ & $11.98^{+1.16}_{-1.18}$ & $11.97^{+0.77}_{-0.74}$\\
    $1.6M_{\odot}$ & $12.11^{+1.33}_{-1.33}$ & $12.03^{+0.98}_{-0.75}$ & $12.05^{+1.32}_{-1.37}$ & $12.00^{+0.90}_{-0.78}$\\
    $2.0M_{\odot}$ & $12.19^{+1.71}_{-1.59}$ & $11.91^{+1.24}_{-1.11}$ & $12.13^{+1.73}_{-1.61}$ & $11.92^{+1.23}_{-1.10}$\\
    \hline
    \end{tabular}
    \end{center}
    Comparison of the 95\% credible interval for the pressure $[\MeV \fmiq]$ and radius [km] of neutron stars when including only HIC experiments and for combined HIC and astrophysics results for two parameterisations of symmetric nuclear matter.
    In particular, we compare the functional form from FOPI used in this work, see Eq.~\eqref{eq:asy_ESNM}, with a general Taylor expansion for symmetric nuclear matter with the same values for the saturation point and the incompressibility but including the third-order parameter $Q=-150\pm 250 \MeV$ at $1\sigma$ using a Gaussian distribution. 
    We find that our results are robust with respect to a variation of this parameterisation and the impact of this choice is at the 5\% level for pressures and 1\% level for radii.
\end{table}

\begin{table}
    \begin{center}
    \renewcommand{\arraystretch}{1.3}
    \caption{{Impact of the proton fraction in $\beta$-equilibrium.}}\label{tab:estimated_free_table}
    \begin{tabular}{|m{0.09\columnwidth}||>{\centering} m{0.165\columnwidth}>{\centering}m{0.165\columnwidth}|>{\centering}m{0.165\columnwidth}>{\centering\arraybackslash}m{0.165\columnwidth}|}
    \hline
    & \multicolumn{2}{c|}{$x_\mathrm{ASY\mbox{-}EOS}$} & \multicolumn{2}{c|}{$0\leq x \leq 0.1$} \\
    $P/R$ & HIC only & Astro+HIC & HIC only & Astro+HIC \\
    \hline\hline
    $1.0 n_{\rm sat}$ & $2.05^{+0.49}_{-0.45}$ & $2.11^{+0.49}_{-0.52}$ & $2.05^{+0.50}_{-0.45}$ & $2.10^{+0.48}_{-0.52}$ \\
    $1.5 n_{\rm sat}$ & $6.06^{+1.85}_{-2.04}$ & $6.25^{+1.90}_{-2.26}$ & $6.02^{+1.89}_{-2.04}$ & $6.23^{+1.81}_{-2.31}$\\
    $2.0 n_{\rm sat}$ & $19.47^{+33.63}_{-11.67}$ & $19.07^{+15.27}_{-10.53}$ & $19.32^{+33.95}_{-11.05}$ & $19.00^{+14.74}_{-10.54}$\\
    $2.5 n_{\rm sat}$ & $47.78^{+75.96}_{-32.96}$ & $45.43^{+40.41}_{-19.11}$ & $48.00^{+78.57}_{-34.40}$ & $45.48^{+39.96}_{-19.28}$\\
    \hline\hline
    $1.0M_{\odot}$ & $11.89^{+0.79}_{-0.98}$ & $11.88^{+0.57}_{-0.76}$ & $11.88^{+0.79}_{-0.98}$ & $11.87^{+0.59}_{-0.75}$\\
    $1.4M_{\odot}$ & $12.06^{+1.13}_{-1.18}$ & $12.01^{+0.78}_{-0.77}$ & $12.05^{+1.14}_{-1.17}$ & $12.00^{+0.77}_{-0.77}$\\
    $1.6M_{\odot}$ & $12.11^{+1.33}_{-1.33}$ & $12.03^{+0.98}_{-0.75}$ & $12.10^{+1.31}_{-1.36}$ & $12.03^{+0.91}_{-0.79}$\\
    $2.0M_{\odot}$ & $12.19^{+1.71}_{-1.59}$ & $11.91^{+1.24}_{-1.11}$ & $12.18^{+1.70}_{-1.61}$ & $11.90^{+1.22}_{-1.14}$\\
    \hline
    \end{tabular}
    \end{center}
    Comparison of the 95\% credible interval for the pressure $[\MeV \fmiq]$ and radius [km] of neutron stars when including only HIC experiments and for combined HIC and astrophysics results for two choices for the proton fraction in $\beta$-equilibrium. 
    For the main results, we compute the proton fraction for the HIC constraints using the EOS functional introduced by the ASY-EOS analysis ($x_\mathrm{ASY\mbox{-}EOS}$). 
    We compare this with a more conservative choice that constrains the proton fraction to be within the range $0\leq x \leq 0.1$ but find only small changes.
\end{table}

\begin{table*}
    \begin{center} 
    \renewcommand{\arraystretch}{1.3}
    \caption{{Impact of the radius constraints for J0740+6220.}}\label{tab:nicer_on_J0740_6220}
    \begin{tabular}{|m{0.09\columnwidth}||>{\centering} m{0.165\columnwidth}|>{\centering} m{0.165\columnwidth}>{\centering}m{0.165\columnwidth}|>{\centering}m{0.165\columnwidth}>{\centering\arraybackslash}m{0.165\columnwidth}|}
    \hline
    & \multicolumn{1}{c|}{}
    & \multicolumn{2}{c|}{Using Ref.\cite{Miller:2021qha,Riley:2021pdl} for J0740+6220} & \multicolumn{2}{c|}{Using Ref.\cite{Fonseca2021} for J0740+6220} \\
    $P/R$ & HIC only & Astro only & Astro+HIC & Astro only & Astro+HIC\\
    \hline\hline
    $1.0 n_{\rm sat}$ & $2.05^{+0.49}_{-0.45}$ & $2.00^{+0.52}_{-0.49}$ & $2.11^{+0.49}_{-0.52}$ & $1.95^{+0.55}_{-0.45}$ & $2.08^{+0.49}_{-0.53}$\\
    $1.5 n_{\rm sat}$ & $6.06^{+1.85}_{-2.04}$ & $5.84^{1.96}_{-2.26}$ & $6.25^{+1.90}_{-2.26}$ & $5.63^{+2.16}_{-2.05}$ & $6.14^{+1.93}_{-2.28}$\\
    $2.0 n_{\rm sat}$ & $19.47^{+33.63}_{-11.67}$ & $18.44^{+16.24}_{-9.69}$ & $19.07^{+15.27}_{-10.53}$ & $17.46^{+15.66}_{-9.27}$ & $18.32^{+14.87}_{-9.60}$\\
    $2.5 n_{\rm sat}$ & $47.78^{+75.96}_{-32.96}$ & $45.05^{+39.80}_{-19.62}$ & $45.43^{+40.41}_{-19.11}$ & $42.23^{+41.75}_{-20.47}$ & $43.22^{+42.66}_{-19.18}$\\
    \hline\hline
    $1.0M_{\odot}$ & $11.89^{+0.79}_{-0.98}$ & $11.76^{+0.65}_{-0.71}$ & $11.88^{+0.57}_{-0.76}$ & $11.68^{+0.71}_{-0.74}$ & $11.82^{+0.68}_{-0.78}$\\
    $1.4M_{\odot}$ & $12.06^{+1.13}_{-1.18}$ & $11.94^{+0.79}_{-0.78}$ & $12.01^{+0.78}_{-0.77}$ & $11.83^{+0.86}_{-0.86}$ & $11.94^{+0.87}_{-0.83}$\\
    $1.6M_{\odot}$ & $12.11^{+1.33}_{-1.33}$ & $11.98^{+0.93}_{-0.79}$ & $12.03^{+0.98}_{-0.75}$ & $11.87^{+1.01}_{-0.93}$ & $11.95^{+1.01}_{-0.91}$\\
    $2.0M_{\odot}$ & $12.19^{+1.71}_{-1.59}$ & $11.88^{+1.23}_{-1.10}$ & $11.91^{+1.24}_{-1.11}$ & $11.74^{+1.44}_{-1.25}$ & $11.77^{+1.42}_{-1.23}$\\
    \hline
    \end{tabular}
    \end{center}
    Comparison of the 95\% credible interval for the pressure $[\MeV \fmiq]$ and radius [km] of neutron stars when including only HIC results, only astrophysical observations, and for combined HIC and astrophysics results when we include the combined mass-radius measurement from NICER~\protect\cite{Miller:2021qha,Riley:2021pdl} or only the radio mass measurement from Ref.\protect\cite{Fonseca2021}. The radius of J0740+6220 estimated by NICER is preferring a stiffer EOS, which agrees well with the constraint from HIC experiments.
\end{table*}

\begin{table}
    \begin{center} 
    \renewcommand{\arraystretch}{1.3}
    \caption{Impact of excluding Danielewicz \textit{et al.}\protect\cite{Dani02Esymm}.}\label{tab:danielewicz}
    \begin{tabular}{|m{0.09\columnwidth}||>{\centering} m{0.165\columnwidth}>{\centering}m{0.165\columnwidth}|>{\centering}m{0.165\columnwidth}>{\centering\arraybackslash}m{0.165\columnwidth}|}
    \hline
    & \multicolumn{2}{c|}{With Danielewicz \textit{et al.}\cite{Dani02Esymm} } & \multicolumn{2}{c|}{Without Danielewicz \textit{et al.}\cite{Dani02Esymm}}\\
    $P/R$ & HIC only & Astro+HIC & HIC only & Astro+HIC\\
    \hline\hline
    $1.0 n_{\rm sat}$ & $2.05^{+0.49}_{-0.45}$ & $2.11^{+0.49}_{-0.52}$ & $2.06^{+0.49}_{-0.45}$ & $2.11^{+0.48}_{-0.52}$\\
    $1.5 n_{\rm sat}$ & $6.06^{+1.85}_{-2.04}$ & $6.25^{+1.90}_{-2.26}$ & $6.08^{+1.83}_{-2.04}$ & $6.25^{+1.89}_{-2.23}$\\
    $2.0 n_{\rm sat}$ & $19.47^{+33.63}_{-11.67}$ & $19.07^{+15.27}_{-10.53}$ & $19.35^{+33.66}_{-10.71}$ & $19.05^{+15.33}_{-10.27}$\\
    $2.5 n_{\rm sat}$ & $47.78^{+75.96}_{-32.96}$ & $45.43^{+40.41}_{-19.11}$ & $47.59^{+79.68}_{-27.46}$ & $45.57^{+40.87}_{-18.89}$\\
    \hline\hline
    $1.0M_{\odot}$ & $11.89^{+0.79}_{-0.98}$ & $11.88^{+0.57}_{-0.76}$ & $11.89^{+0.79}_{-0.98}$ & $11.88^{+0.56}_{-0.78}$\\
    $1.4M_{\odot}$ & $12.06^{+1.13}_{-1.18}$ & $12.01^{+0.78}_{-0.77}$ & $12.06^{+1.12}_{-1.19}$ & $12.01^{+0.78}_{-0.77}$\\
    $1.6M_{\odot}$ & $12.11^{+1.33}_{-1.33}$ & $12.03^{+0.98}_{-0.75}$ & $12.11^{+1.32}_{-1.34}$ & $12.03^{+0.92}_{-0.80}$\\
    $2.0M_{\odot}$ & $12.19^{+1.71}_{-1.59}$ & $11.91^{+1.24}_{-1.11}$ & $12.18^{+1.70}_{-1.61}$ & $11.91^{+1.17}_{-1.15}$\\
    \hline
    \end{tabular}
    \end{center}
    Comparison of the 95\% credible interval for the pressure $[\MeV \fmiq]$ and radius [km] of neutron stars when including only HIC experiments and for combined HIC and astrophysics results with and without the inclusion of the constraint from Danielewicz \textit{et al.}\protect\cite{Dani02Esymm}. By comparing the HIC-only results, we conclude that the constraint from Danielewicz \textit{et al.}\protect\cite{Dani02Esymm} has a small impact on our study.
\end{table}

%\FloatBarrier
%\bibliographysuppl{refs}
%\bibliographystylesuppl{naturemag}

\end{document}